\pdfoutput=1
\documentclass[letterpaper,twoside,11pt]{article}

\usepackage{lipsum} 

\usepackage[sc]{mathpazo} 
\usepackage[T1]{fontenc} 
\usepackage{microtype} 

\usepackage[letterpaper,hmarginratio=1:1,top=32mm,left=20mm,right=20mm,columnsep=20pt]{geometry} 
\usepackage[hang, small,labelfont=bf,up,textfont=it,up]{caption} 
\usepackage{booktabs} 
\usepackage{float} 
\usepackage{hyperref} 

\usepackage{lettrine} 
\usepackage{paralist} 

\usepackage{abstract} 

\usepackage{titlesec} 
\titleformat{\section}[block]{\large\scshape\centering}{\thesection.}{1em}{} 
\titleformat{\subsection}[block]{\large}{\thesubsection.}{1em}{} 

\usepackage{fancyhdr} 
\usepackage{indentfirst}

    \usepackage[utf8]{inputenc}                         
    \usepackage[english]{babel}               
    \usepackage{amsmath}                                
    \usepackage{graphicx}                               
    \usepackage{tikz}
    \usepackage{caption,subcaption}                     

    \usepackage{appendix}                               
    \usepackage{pdfpages}                               
    \usepackage{array}                                  
    \usepackage{algorithm}
    \usepackage[noend]{algpseudocode}
    \usepackage{eqparbox}
    \makeatletter
    \def\BState{\State\hskip-\ALG@thistlm}
    \makeatother
    \usepackage{mathtools}
    \usepackage{amsthm}
    
    \usepackage{enumerate}                              
    \usepackage{sectsty}                                
    \usepackage{todonotes}                              
    \usepackage[space]{grffile}                         
    \usepackage{siunitx}                               

    \numberwithin{equation}{section} 
    \numberwithin{figure}{section} 
    \numberwithin{table}{section} 

    \renewcommand{\labelitemii}{\leavevmode \hbox to
    1.2ex {\hss \vrule height .9ex width .7ex depth -.2ex\hss }} 

    \DeclarePairedDelimiter\paren{(}{)}%
    \DeclarePairedDelimiter\bra{[}{]}%
    \DeclarePairedDelimiter\curly{\{}{\}}%
    \DeclarePairedDelimiter\abs{\lvert}{\rvert}%
    \DeclarePairedDelimiter\norm{\lVert}{\rVert}%

    \usetikzlibrary{shapes,arrows,fit,calc,positioning,automata,decorations.markings}

    \tikzstyle{block} = [draw, fill=white, rectangle,
        minimum height=2.5em, minimum width=2.5em]
    \tikzstyle{sum} = [draw, fill=white, circle, node distance=1cm]
    \tikzstyle{input} = [coordinate]
    \tikzstyle{output} = [coordinate]
    \tikzstyle{pinstyle} = [pin edge={to-,thin,black}]

    \tikzset{XOR/.style={draw,circle,append after command={
            [shorten >=\pgflinewidth, shorten <=\pgflinewidth,]
            (\tikzlastnode.north) edge (\tikzlastnode.south)
            (\tikzlastnode.east) edge (\tikzlastnode.west)
            }
        }
    }
    \tikzset{line/.style={draw, -latex'}}
    \tikzset{mdec/.style={decoration={markings, mark=at position 0.5 with {\fill[black!80!white] circle (1.5pt);}},postaction={decorate}}}

    \usepackage{listings}                               
    \usepackage{courier}                                

    \AtBeginDocument{\numberwithin{lstlisting}{section}}  
    
    \definecolor{background}{gray}{.98}                 
    \definecolor{comments}{RGB}{51,102,0}               
    \definecolor{keywords}{RGB}{0,0,120}                
    \definecolor{keywords2}{RGB}{204,0,102}             
    \definecolor{numbers}{RGB}{127, 0, 127}             

    \definecolor{Maroon}{RGB}{128, 0, 0}

    \lstdefinelanguage{ICAI-RiSC-16}
   {morekeywords={add,sub,mul,mulu,nand,sll,sra,srl,sltu,addi,sw,lw,beq,jalr,lui,la,med},
        sensitive=false,
        morecomment=[l]{\#},
        morecomment=[l][\color{Maroon}]{.},
        morekeywords=[2]{r0, r1, r2, r3, r4, r5, r6, r7},
        keywordstyle=[2]{\color{violet}},
    }

    \lstdefinestyle{customvhdl}{
        language=vhdl,                              
        morekeywords = {},
        morestring=[b][\color{numbers}]',
        morestring=[b][\color{numbers}]",
        emph={std_logic,std_logic_vector,integer,unsigned,to_unsigned},
        emphstyle={\color{keywords2}\bfseries},%
    }

    \lstdefinestyle{customcpp}{
        language = C++,
    }

\usepackage{upgreek}
\usepackage{ amssymb }
\usepackage{relsize}
\usepackage{float}
\newfloat{algorithm}{ht}{lop}

\usepackage{setspace}
\let\Algorithm\algorithm
\renewcommand\algorithm[1][]{\Algorithm[#1]\setstretch{1.4}}

\newif\ifEXTENDED
\EXTENDEDtrue

\newcommand{\ham}[1]{\mathrm{H}\bra{#1}}
\newcommand{\Ham}[1]{\mathrm{H}\bra*{#1}}
\newcommand{\Hams}[1]{\mathrm{H}^{\ast}_{\epsilon}\paren*{#1}}
\newcommand{\Sum}{\sum}

\title{\vspace{-15mm}\parbox{0.75\textwidth}{\centering\fontsize{24pt}{10pt}\selectfont\textbf{A Simple Power Analysis Attack on the Twofish Key Schedule}}} 

\author{
\large
\textsc{Jose Javier Gonzalez Ortiz}\thanks{\href{mailto:jjgo@umich.edu}{jjgo@umich.edu}}\\[2mm] 
\normalsize University of Michigan  \\ \and \textsc{Kevin J. Compton\thanks{\href{mailto:kjc@umich.edu}{kjc@umich.edu}}}\\[2mm] 
\normalsize University of Michigan \\ 
}
\date{\today}

\begin{document}

\maketitle 


\vspace{-.5cm}
\begin{abstract}

\noindent This paper introduces an SPA power attack on the 8-bit implementation of the Twofish block cipher. The attack is able to unequivocally recover the secret key even under substantial amounts of error. An initial algorithm is described using exhaustive search on error free data. An error resistant algorithm is later described. It employs several threshold preprocessing stages followed by a combined approach of least mean squares and an optimized Hamming mask search. Further analysis of 32 and 64-bit Twofish implementations reveals that they are similarly vulnerable to the described SPA attack.

\end{abstract}


\section{Introduction}
In 1997, the National Institute of Standards and Technology (NIST) started the Advanced Encryption Standard process in order to designate a successor of the aged Digital Encryption Standard (DES). Among the five finalists is the Twofish block cipher, and although in the end Rijndael was designated as the winner of the AES contest, Twofish was one of the strongest contenders, excelling in categories such as general security, hardware performance and design features. After the introduction of linear and differential attacks they became a concern in the design of new ciphers. Thus, the encryption and key schedule algorithms of the submissions were designed to prevent these types of attacks. Nevertheless, among the finalists both AES and Serpent have been found to be susceptible to \emph{side-channel attacks}. These attacks focus on the information that can be gained from the physical implementation of cryptosystems in especially accessible systems such as smart cards.

In 1999 Biham and Shamir \cite{biham1999power} carried out a preliminary analysis of power attack susceptibility of various AES candidates. They assessed that Twofish Key Schedule had a complex structure and seemed not to reveal direct information on the key bits from Hamming measurements. The results presented in this work refute this assertion since an efficient and robust SPA attack that retrieves the secret key was found.

We present a side channel attack on the Twofish key schedule which finds the secret key in just one execution of the algorithm. Further analysis of the algorithm and the different existing implementations render all of them vulnerable to this attack, posing a concern for the way 8-bit manipulations are described and performed in the Twofish Key Schedule.

We will start by providing a background in the Twofish Block Cipher operation and the formulation of its associated Key Schedule in Section \ref{sec:twofish_block_cipher}. Section \ref{sec:twofish_key_schedule_power_analysis_attack} describes an attack provided error free data by employing a reduced exhaustive search for each byte of the key. The attack is then incrementally improved in Section \ref{sec:attack_in_the_presence_of_error} to cope with a more than significant amount of measurement error. Section \ref{sec:results} displays the accuracies achieved under varying degrees of error and key size as well as the elapsed times. These results are later analyzed on Section \ref{sec:discussion} and further work is outlined in Section \ref{sec:conclusion}.

\subsection{Previous Work} 
\label{sub:previous_work}

Side Channel Attacks were first described in 1996 by Kocher \cite{kocher1996timing} with the introduction of timing attacks, which had the ability of compromising a cryptosystem by analyzing the time taken to execute specific parts of cryptographic algorithms. However, these attacks can be easily overcome by employing simple software countermeasures.
Since access to the hardware was one of the assumptions, attacks that used the power consumption of the cryptosystem started developing. The first research power attack is credited to \emph{Kocher et al} \cite{kocher1999differential}, in which they were able to obtain information of the data manipulated by the processor by carefully measuring the power consumption of a CMOS chip. In his work, two different types of power attacks are described:
\begin{itemize}
  \item \textbf{Differential Power Attacks} (DPA) -- In an analogous fashion to classical differential cryptanalysis, DPAs try to find patterns and relationships between plaintexts and their associated power traces. Similarly, a large amount of samples are required for the results to convey statistical significance.
  \item \textbf{Simple Power Attacks} (SPA) -- An SPA focuses in particular vulnerabilities of the algorithm design. These vulnerabilities  could leak enough sensitive  information to compromise the confidentiality of the encryption or even the secret key itself.
\end{itemize}

Since microprocessors perform discrete operations on blocks of data in a sequential fashion, physical imperfections of the system make possible to correlate the Hamming weights of the manipulated values and the power utilization. Research has proved that this correlation is significant enough to find Hamming weights of numerous intermediate variables from the power utilization, as shown in the works of \cite{Messerges:2002:ESS:570513.570522} and \cite{Mayer-Sommer01smartlyanalyzing}.

The concern of these types of attacks has lead to a number of research efforts \cite{Messerges:2000:SAF:647935.740925}, \cite{Mangard:2007:PAA:1208234} to thwart them by masquerading the power consumption values in order to break the correlation that Power Attack use as a basis to acquire information. Nevertheless, they are still far from being industry standards and most modern smartcards and ASICs do not yet include mechanisms like those ones by default. Thus, for the purpose of this paper said techniques have not been considered.

Recent research efforts have shown that both Rijndael \cite{VanLaven2005SideCA} and Serpent \cite{Compton2009ASP} key schedules are susceptible to Simple Power Attacks. Both attacks used a power trace of the algorithm to unequivocally recover the secret key used in the encryption. In both cases the weakness arose from the key schedule computation; patterns in the hamming weights of the key schedule algorithm revealed enough information to compromise the secret key. Twofish's key schedule follows different principles and constructions to generate the necessary subkeys but it is nonetheless susceptible to this type of attack as this paper demonstrates.

Even though Twofish smart-card implementation performance  and versatility have been thoroughly analyzed \cite{Keating99performanceanalysis}, \cite{Rizvi:2011:PAA:2013882.2014256}, no power attacks have been found for the encryption algorithm or the key schedule. Nevertheless, analysis on the key schedule such as \cite{Mirza99anobservation} have outlined some deficiencies and weaknesses in the cipher.



\section{Twofish Block Cipher} 
\label{sec:twofish_block_cipher}

Twofish is a symmetric key block cipher with a block size of 128 bits and key sizes up to 256 bits. It was one of the five finalists of the Advanced Encryption Standard contest. It was submitted by Schneier et al. \cite{Schneier98twofish:a}.

Twofish features pre-computed key-dependent S-boxes, and a relatively complex key schedule. One half of an n-bit key is used as the actual encryption key and the other half of the n-bit key is used to modify the encryption algorithm. Twofish borrows some elements from other designs such as a Feistel structure from DES or the pseudo-Hadamard transform \cite{surhone2010pseudo} from the SAFER family of ciphers \cite{Massey94saferk-64:}.

This section will briefly introduce the encryption scheme and key schedule in the Twofish block cipher following the notation and terminology from \cite{Schneier98twofish:a}.

\subsection{The Twofish Encryption Algorithm} 
\label{sub:encryption_algorithm}

The Twofish encryption is a 16 round Feistel Network with both input and output whitening. Each round operates only in the higher 64 bits of the block and swaps both halves. A total of 40 subkeys are generated from the secret key, with each key being 32 bits. Keys $K_0\ldots K_3$ are used for the input whitening, $K_4\ldots K_7$ are used for the output whitening and $K_8\ldots K_{39}$ are used as the round subkeys. Each round employs a function $F$ which is a key-dependent permutation on 64-bit values.  The function $F$ splits the 64-bit string into two 32-bit substrings and applies the $g$ function to each half. The function $g$ applies a fixed number of S-box substitutions and XORs with parts of the secret key (the number of steps this is performed depends on the size of the key).

\begin{figure}
    \includegraphics[width=0.9\textwidth]{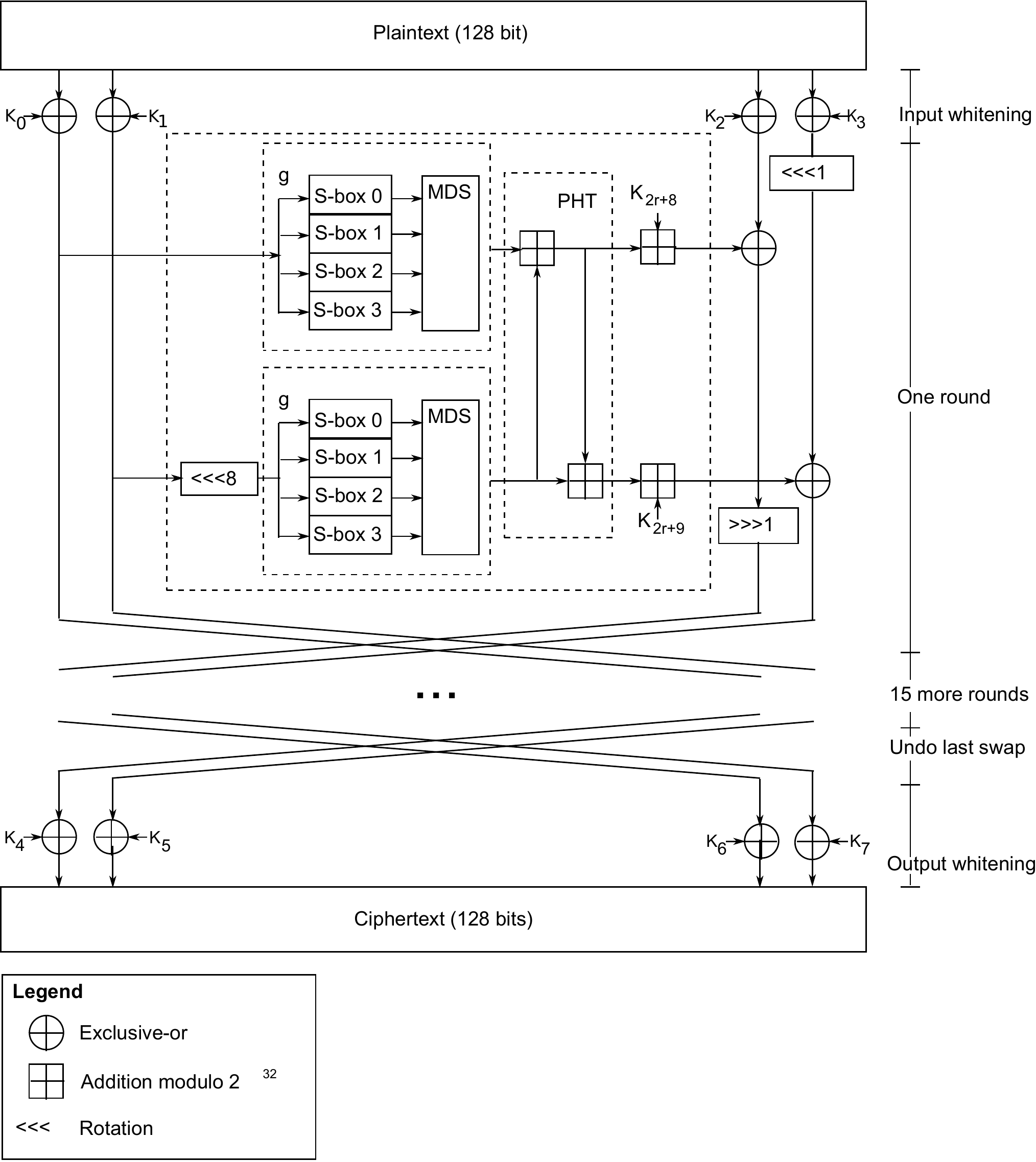}
    \centering
    \caption{Twofish algorithm}
    \label{fig:twofishalgo}
\end{figure}

This is followed by a MDS (Maximum Distance Separable) Matrix transform, a Pseudo-Hadamard Transform and round key XOR with the two subkeys associated for that said round. Therefore each round $r$ uses two subkeys as round key, namely $K_{2r+8}$ and $K_{2r+9}$. Finally, the output is XORed with one half of the block following the Feistel Network scheme. Bit wise rotations and shifts are performed at strategic points in the encryption to maximize diffusion. They have been omitted to simplify the explanation. The algorithm can be visualized in Figure \ref{fig:twofishalgo}

\subsection{The Twofish Key Schedule} 
\label{sub:key_schedule}

The key schedule has to provide 40 32-bit words of expanded key $K_{0} ,\ldots, K_{39}$. Twofish is defined for sizes $N = \curly*{128,192,256}$. Keys shorter can be padded with zeros to the next larger key length. The key is split into $2R$ 32-bit words $K = \paren*{M_{0},M_{1},\ldots M_{2R-1}}$ where $R = N/64$. To generate each sub-key all the bits in the secret key are employed.
The secret key is also employed to derive the vector $S = (S_{R-1}, S_{R-2},\ldots,S_0)$ which is obtained by applying a Reed Solomon Transformation to the key. These values are used in the $g$ function inside the $F$ permutation as part of the encryption algorithm.

\begin{figure}
    \centering
    \includegraphics[width=0.85\textwidth]{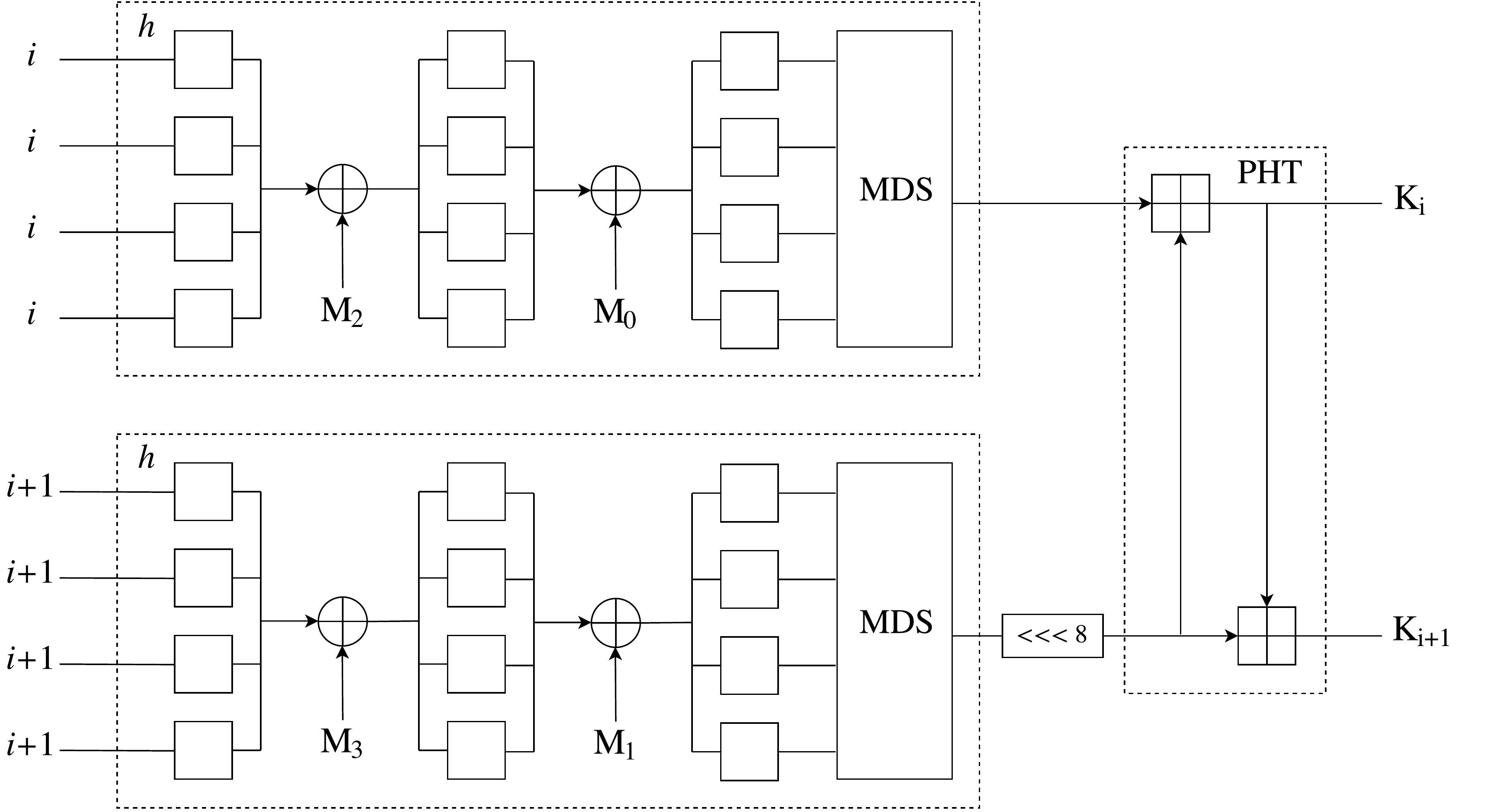}
    \vspace{.5cm}
    \caption{Twofish Key Schedule (128-bit key)}
    \label{fig:twofishkey}
\end{figure}

Subkeys are generated in even-odd pairs $K_{i}, K_{i+1}$, with even $i$. To generate this pair of keys, two 32-words are initialized, the first word has all its bytes equal to $i$ and the second word has all its bytes equal to $i+1$. Then, both words go through $R$ rounds, each round being composed of a specific substitution box arrangement followed by an XOR with a corresponding part of the secret key. This combination makes the S-boxes key dependent. For example, for $\abs*{K} = 128 $ we have $R = 2$, so the $h$ function will have two rounds as shown in Figure \ref{fig:twofishkey}.

Next, the words go through another S-box substitution, and through a MDS transform. All of these 32 bit S-boxes are composed of a predefined choice of two 8-bit permutations $q0$ and $q1$. This choice is fixed and only depends on the stage of the function we are in. The transformation up to this point is defined to be the $h$ function. Next, an 8 bit right rotation is applied in the odd word. Finally a Pseudo Hadamard Transform is applied to both values resulting in the pair of subkeys $K_{2n}, K_{2n+1}$.

We can see that the procedure is quite similar to that of the round function and in fact the $g$ function can be expressed in terms of the $h$ function. However,  the round function does not use directly the bits of the secret key, but instead a a Reed Solomon Matrix is applied beforehand as we previously mentioned. For a further explanation refer to the equations and diagrams in \cite{Schneier98twofish:a}.


\subsection{Twofish Sub-key generation} 
\label{sub:subsection_name}

The main problem when using the notation in \cite{Schneier98twofish:a} is that it describes the algorithm mainly in operations of 32-bit words. Since we would like to perform an attack that will use the trace of Hamming weights, we are interested in having a mathematical formulation that operates on 8-bit values. Therefore in this section we will derive the equations for the Twofish key schedule in byte form, laying out the mathematical notation that we will use in the Simple Power Attack in the next section.

Although the secret key values are both used for the round function and the key schedule, the attack was found just by looking at Hamming weights of the key schedule algorithm, so only notation for the $h$ function is going to be introduced.

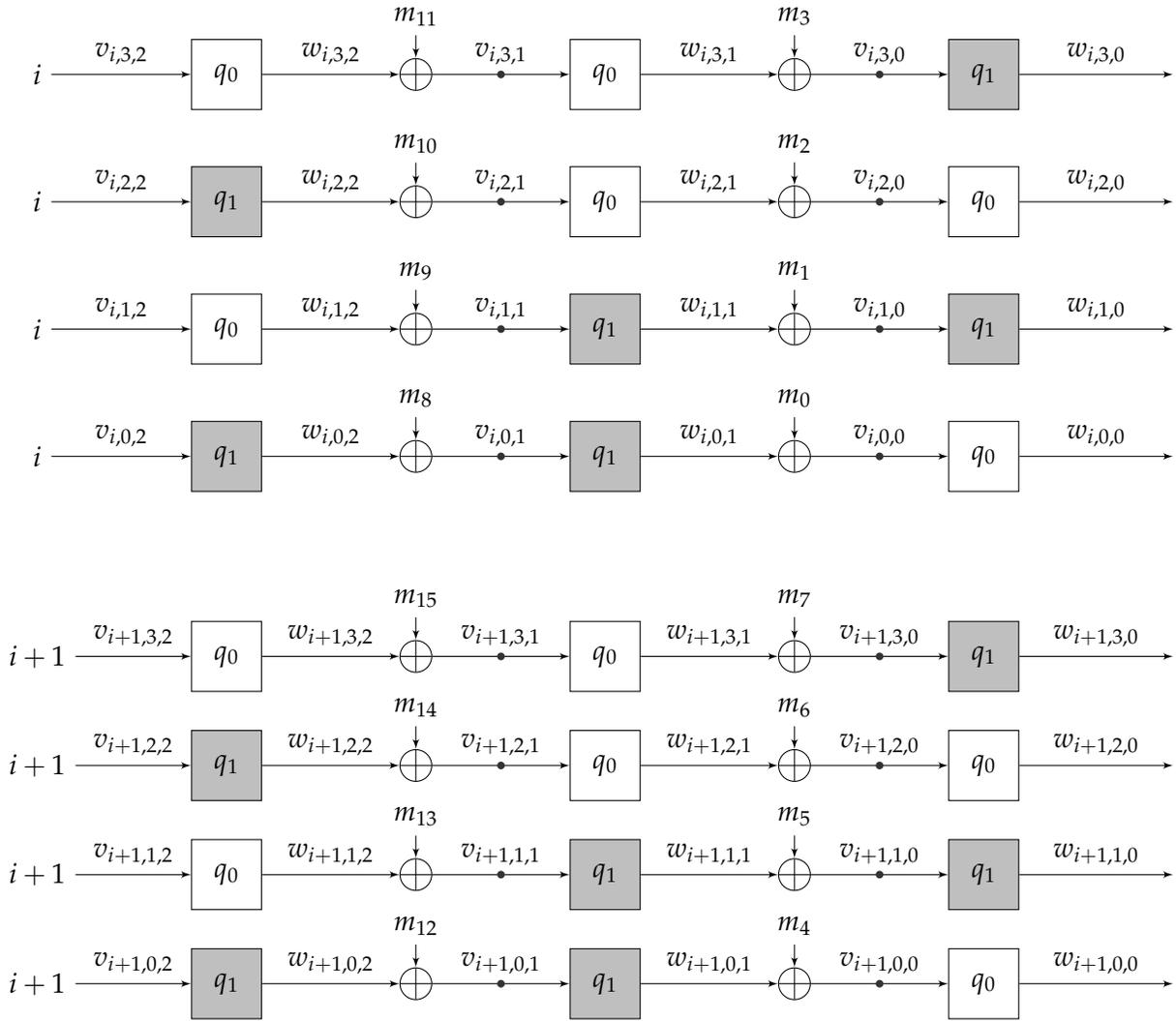
\begin{figure}[!ht]
\centering

\begin{tikzpicture}[auto,node distance=2.6cm,>=latex']

    \node at (0,0) (XOR-32) {$i$};

    \node [block, right of=XOR-32] (Q-32) {$q_0$};
    \node [XOR, right of=Q-32,scale=1.2] (XOR-31) {};
    \node [block, right of=XOR-31] (Q-31) {$q_0$};
    \node [XOR, right of=Q-31,scale=1.2] (XOR-30) {};
    \node [block, right of=XOR-30, fill=gray!50!white] (Q-30) {$q_1$};

    \node [output, right of=Q-30] (out-3) {};

    \node [above of =XOR-31, node distance=.8cm] (m-31) {$m_{11}$};
    \node [above of =XOR-30, node distance=.8cm] (m-30) {$m_{3}$};

    \draw [->] (XOR-32) -- node {$v_{i,3,2}$} (Q-32);
    \draw [->] (Q-32) -- node {$w_{i,3,2}$} (XOR-31);
    \draw [->] (m-31) -- node {} (XOR-31);

    \draw [->,mdec] (XOR-31) -- node {$v_{i,3,1}$} (Q-31);
    \draw [->] (Q-31) -- node {$w_{i,3,1}$} (XOR-30);
    \draw [->] (m-30) -- node {} (XOR-30);

    \draw [->,mdec] (XOR-30) -- node {$v_{i,3,0}$} (Q-30);
    \draw [->] (Q-30) -- node {$w_{i,3,0}$} (out-3);

    \node at (0,-1.75) (XOR-22) {$i$};

    \node [block, right of=XOR-22, fill=gray!50!white] (Q-22) {$q_1$};
    \node [XOR, right of=Q-22,scale=1.2] (XOR-21) {};
    \node [block, right of=XOR-21] (Q-21) {$q_0$};
    \node [XOR, right of=Q-21,scale=1.2] (XOR-20) {};
    \node [block, right of=XOR-20] (Q-20) {$q_0$};

    \node [output, right of=Q-20] (out-2) {};

    \node [above of =XOR-21, node distance=.8cm] (m-21) {$m_{10}$};
    \node [above of =XOR-20, node distance=.8cm] (m-20) {$m_{2}$};

    \draw [->] (XOR-22) -- node {$v_{i,2,2}$} (Q-22);
    \draw [->] (Q-22) -- node {$w_{i,2,2}$} (XOR-21);
    \draw [->] (m-21) -- node {} (XOR-21);

    \draw [->,mdec] (XOR-21) -- node {$v_{i,2,1}$} (Q-21);
    \draw [->] (Q-21) -- node {$w_{i,2,1}$} (XOR-20);
    \draw [->] (m-20) -- node {} (XOR-20);

    \draw [->,mdec] (XOR-20) -- node {$v_{i,2,0}$} (Q-20);
    \draw [->] (Q-20) -- node {$w_{i,2,0}$} (out-2);

    \node at (0,-3.5) (XOR-12) {$i$};

    \node [block, right of=XOR-12] (Q-12) {$q_0$};
    \node [XOR, right of=Q-12,scale=1.2] (XOR-11) {};
    \node [block, right of=XOR-11, fill=gray!50!white] (Q-11) {$q_1$};
    \node [XOR, right of=Q-11,scale=1.2] (XOR-10) {};
    \node [block, right of=XOR-10, fill=gray!50!white] (Q-10) {$q_1$};

    \node [output, right of=Q-10] (out-1) {};

    \node [above of =XOR-11, node distance=.8cm] (m-11) {$m_{9}$};
    \node [above of =XOR-10, node distance=.8cm] (m-10) {$m_{1}$};

    \draw [->] (XOR-12) -- node {$v_{i,1,2}$} (Q-12);
    \draw [->] (Q-12) -- node {$w_{i,1,2}$} (XOR-11);
    \draw [->] (m-11) -- node {} (XOR-11);

    \draw [->,mdec] (XOR-11) -- node {$v_{i,1,1}$} (Q-11);
    \draw [->] (Q-11) -- node {$w_{i,1,1}$} (XOR-10);
    \draw [->] (m-10) -- node {} (XOR-10);

    \draw [->,mdec] (XOR-10) -- node {$v_{i,1,0}$} (Q-10);
    \draw [->] (Q-10) -- node {$w_{i,1,0}$} (out-1);

    \node at (0,-5.25) (XOR-02) {$i$};

    \node [block, right of=XOR-02, fill=gray!50!white] (Q-02) {$q_1$};
    \node [XOR, right of=Q-02,scale=1.2] (XOR-01) {};
    \node [block, right of=XOR-01, fill=gray!50!white] (Q-01) {$q_1$};
    \node [XOR, right of=Q-01,scale=1.2] (XOR-00) {};
    \node [block, right of=XOR-00] (Q-00) {$q_0$};

    \node [output, right of=Q-00] (out-0) {};

    \node [above of =XOR-01, node distance=.8cm] (m-01) {$m_{8}$};
    \node [above of =XOR-00, node distance=.8cm] (m-00) {$m_{0}$};

    \draw [->] (XOR-02) -- node {$v_{i,0,2}$} (Q-02);
    \draw [->] (Q-02) -- node {$w_{i,0,2}$} (XOR-01);
    \draw [->] (m-01) -- node {} (XOR-01);

    \draw [->,mdec] (XOR-01) -- node {$v_{i,0,1}$} (Q-01);
    \draw [->] (Q-01) -- node {$w_{i,0,1}$} (XOR-00);
    \draw [->] (m-00) -- node {} (XOR-00);

    \draw [->,mdec] (XOR-00) -- node {$v_{i,0,0}$} (Q-00);
    \draw [->] (Q-00) -- node {$w_{i,0,0}$} (out-0);

    
    \node at (0,-8) (XOR-32b) {$i+1$};

    \node [block, right of=XOR-32b] (Q-32b) {$q_0$};
    \node [XOR, right of=Q-32b,scale=1.2] (XOR-31b) {};
    \node [block, right of=XOR-31b] (Q-31b) {$q_0$};
    \node [XOR, right of=Q-31b,scale=1.2] (XOR-30b) {};
    \node [block, right of=XOR-30b, fill=gray!50!white] (Q-30b) {$q_1$};

    \node [output, right of=Q-30b] (out-3b) {};

    \node [above of =XOR-31b, node distance=.8cm] (m-31b) {$m_{15}$};
    \node [above of =XOR-30b, node distance=.8cm] (m-30b) {$m_{7}$};

    \draw [->] (XOR-32b) -- node {$v_{i+1,3,2}$} (Q-32b);
    \draw [->] (Q-32b) -- node {$w_{i+1,3,2}$} (XOR-31b);
    \draw [->] (m-31b) -- node {} (XOR-31b);

    \draw [->,mdec] (XOR-31b) -- node {$v_{i+1,3,1}$} (Q-31b);
    \draw [->] (Q-31b) -- node {$w_{i+1,3,1}$} (XOR-30b);
    \draw [->] (m-30b) -- node {} (XOR-30b);

    \draw [->,mdec] (XOR-30b) -- node {$v_{i+1,3,0}$} (Q-30b);
    \draw [->] (Q-30b) -- node {$w_{i+1,3,0}$} (out-3b);

    \node at (0,-9.5) (XOR-22b) {$i+1$};

    \node [block, right of=XOR-22b, fill=gray!50!white] (Q-22b) {$q_1$};
    \node [XOR, right of=Q-22b,scale=1.2] (XOR-21b) {};
    \node [block, right of=XOR-21b] (Q-21b) {$q_0$};
    \node [XOR, right of=Q-21b,scale=1.2] (XOR-20b) {};
    \node [block, right of=XOR-20b] (Q-20b) {$q_0$};

    \node [output, right of=Q-20b] (out-2b) {};

    \node [above of =XOR-21b, node distance=.8cm] (m-21b) {$m_{14}$};
    \node [above of =XOR-20b, node distance=.8cm] (m-20b) {$m_{6}$};

    \draw [->] (XOR-22b) -- node {$v_{i+1,2,2}$} (Q-22b);
    \draw [->] (Q-22b) -- node {$w_{i+1,2,2}$} (XOR-21b);
    \draw [->] (m-21b) -- node {} (XOR-21b);

    \draw [->,mdec] (XOR-21b) -- node {$v_{i+1,2,1}$} (Q-21b);
    \draw [->] (Q-21b) -- node {$w_{i+1,2,1}$} (XOR-20b);
    \draw [->] (m-20b) -- node {} (XOR-20b);

    \draw [->,mdec] (XOR-20b) -- node {$v_{i+1,2,0}$} (Q-20b);
    \draw [->] (Q-20b) -- node {$w_{i+1,2,0}$} (out-2b);

    \node at (0,-11) (XOR-12b) {$i+1$};

    \node [block, right of=XOR-12b] (Q-12b) {$q_0$};
    \node [XOR, right of=Q-12b,scale=1.2] (XOR-11b) {};
    \node [block, right of=XOR-11b, fill=gray!50!white] (Q-11b) {$q_1$};
    \node [XOR, right of=Q-11b,scale=1.2] (XOR-10b) {};
    \node [block, right of=XOR-10b, fill=gray!50!white] (Q-10b) {$q_1$};

    \node [output, right of=Q-10b] (out-1b) {};

    \node [above of =XOR-11b, node distance=.8cm] (m-11b) {$m_{13}$};
    \node [above of =XOR-10b, node distance=.8cm] (m-10b) {$m_{5}$};

    \draw [->] (XOR-12b) -- node {$v_{i+1,1,2}$} (Q-12b);
    \draw [->] (Q-12b) -- node {$w_{i+1,1,2}$} (XOR-11b);
    \draw [->] (m-11b) -- node {} (XOR-11b);

    \draw [->,mdec] (XOR-11b) -- node {$v_{i+1,1,1}$} (Q-11b);
    \draw [->] (Q-11b) -- node {$w_{i+1,1,1}$} (XOR-10b);
    \draw [->] (m-10b) -- node {} (XOR-10b);

    \draw [->,mdec] (XOR-10b) -- node {$v_{i+1,1,0}$} (Q-10b);
    \draw [->] (Q-10b) -- node {$w_{i+1,1,0}$} (out-1b);

    \node at (0,-12.5) (XOR-02b) {$i+1$};

    \node [block, right of=XOR-02b, fill=gray!50!white] (Q-02b) {$q_1$};
    \node [XOR, right of=Q-02b,scale=1.2] (XOR-01b) {};
    \node [block, right of=XOR-01b, fill=gray!50!white] (Q-01b) {$q_1$};
    \node [XOR, right of=Q-01b,scale=1.2] (XOR-00b) {};
    \node [block, right of=XOR-00b] (Q-00b) {$q_0$};

    \node [output, right of=Q-00b] (out-0b) {};

    \node [above of =XOR-01b, node distance=.8cm] (m-01b) {$m_{12}$};
    \node [above of =XOR-00b, node distance=.8cm] (m-00b) {$m_{4}$};

    \draw [->] (XOR-02b) -- node {$v_{i+1,0,2}$} (Q-02b);
    \draw [->] (Q-02b) -- node {$w_{i+1,0,2}$} (XOR-01b);
    \draw [->] (m-01b) -- node {} (XOR-01b);

    \draw [->,mdec] (XOR-01b) -- node {$v_{i+1,0,1}$} (Q-01b);
    \draw [->] (Q-01b) -- node {$w_{i+1,0,1}$} (XOR-00b);
    \draw [->] (m-00b) -- node {} (XOR-00b);

    \draw [->,mdec] (XOR-00b) -- node {$v_{i+1,0,0}$} (Q-00b);
    \draw [->] (Q-00b) -- node {$w_{i+1,0,0}$} (out-0b);

\end{tikzpicture}
\caption{Key schedule computations for a 128-bit key}
\label{fig:keydetail}
\end{figure}

In section \ref{sub:key_schedule}  we described the $h$ function involved in the key schedule procedure, which is depicted in Figure \ref{fig:twofishkey} for a 128-bit key. Expanding the diagram to take into account all the descriptions of the key schedule, we get the layout shown in Figure \ref{fig:keydetail}.

Here $i$ is an even integer and the bytes $m_0,\ldots,m_{15}$ are the 16 bytes of the key.  Blocks $q0$ and $q1$ are byte substitution boxes implemented using look-up tables, and their position along the rows and rounds is part of the specification of the algorithm. Variables $v$ and $w$ will be later used to express the relationship between the different rounds of the algorithm.Finally, the grey dots represent the places where a Hamming weight is retrieved from the trace in the algorithm described in Section \ref{sec:twofish_key_schedule_power_analysis_attack}.

As we can see from the diagram we will need at least three indexes in order to identify a particular Hamming weight inside the $h$ function structure.
\begin{description}
  \item[${i}$] -- \;\;Identifies the sub-key we are generating as stated previously. In general $i$ will range from $0$ to $39$, since 40 subkeys need to be generated.
  \item[${j}$] -- \;\;Specifies the byte within the 32-bit word taken in the $h$ function. $j$ will range from $0$ to $3$. In the diagram this is represented as rows.
  \item[${k}$] -- \;Will identify in which round inside the $h$ function we are at. Since the number of rounds will depend on the size of the secret key, $k$ will range from $0$ to $R$, where $R = \abs{K}/64$.
\end{description}

In general, the key consists of $8R$ bytes, which are all used in every execution of the $h$ function.
\begin{equation}
  K = \curly*{m_{0},m_{1},\ldots,m_{8R-1}}
\end{equation}

In order to describe the relationships between the intermediate values within the $h$ function we can use two byte vectors $\boldsymbol{V}$ and $\boldsymbol{W}$ to represent the values before and after applying a S-box respectively.

\begin{align}
\boldsymbol{V} &= \curly*{v_{i\,j\,k} \phantom{.}\in \curly*{0,1}^{8} : \;\;i= 0,1,\ldots,39 \quad j=0,1,2,3 \quad k = 0,1,\ldots,R}\\
\boldsymbol{W} &= \curly*{w_{i\,j\,k} \in \curly*{0,1}^{8} : \;\;i= 0,1,\ldots,39 \quad j=0,1,2,3 \quad k = 0,1,\ldots,R}
\end{align}

This notation is displayed in Figure \ref{fig:keydetail}. One important observation is that the index $k$ decreases as we apply more rounds. This comes from the fact that increasing the key size adds more rounds to the left of the $h$ function, leaving the rest of the rounds unaltered. Since the layout of $q0$ and $q1$ boxes depends both on $j$ and $k$, the simplest way to define this relationship is to increment $k$ from right to left. Furthermore, to generalize the notion of the S-boxes to these indexes, we can create a matrix $P$ whose elements $P_{jk}$ are 0 when the substitution used in row $j$ and round $k$ is $q0$ and $1$ if it is $q1$. This allows us to succinctly store the disposition of $q0$ and $q1$ in the algorithm and define a permutation $Q_{jk}[x]$ that will evaluate to either $q0$ or $q1$.

\begin{equation}
Q_{jk}[x] = \begin{cases}
q_{0}[x] \quad \text{if} \quad P_{jk} = 0\\
q_{1}[x] \quad \text{if} \quad P_{jk} = 1
\end{cases}\qquad\qquad
  P = \begin{pmatrix}
  0 & 1 & 1 & 0 & 1 \\
  1 & 1 & 0 & 0 & 0 \\
  0 & 0 & 1 & 1 & 0 \\
  1 & 0 & 0 & 1 & 1 \\
  \end{pmatrix}
\end{equation}

  

Given all these definitions we can expresses the function $h$ using the following relations.
\begingroup
\addtolength{\jot}{1em}
\begin{align}
  v_{i,j,R} &= i \label{eq:init}\\
  w_{i,j,k} &= Q_{jk}[v_{i,j,k}] \label{eq:wQv}\\
  v_{i,j,(k-1)} &= w_{i,j,k} \oplus m_{l} \label{eq:vwm}
\end{align}
\endgroup

As we see in equation \eqref{eq:vwm}, a index $l$ for the specific byte of the key is required. In the diagram we could see directly where every byte of the key was used. In general, half of the executions of $h$ will require even words of the key $M_{e} = \curly*{M_{0},M_{2},\ldots}$ and the other half  will require odd words of the key $M_{0} = \curly*{M_{1},M_{3}\ldots}$.

We can express $l$ in terms of the rest of the indexes, so from now on when $l$ is used, it can be thought as a function of $i,j,k$.

\begin{equation}
  l(i,j,k) = 8(k-1)+j+4(i\bmod 2)
\end{equation}


\section{Twofish Key Schedule Power Analysis Attack} 
\label{sec:twofish_key_schedule_power_analysis_attack}

From the formulation described in the previous section we can see that each byte of the key $m_{l}$ is used $20$ different times in order to generate the $40$ round subkeys. Since $m_{l} \in \curly*{0,1}^{8}$, knowing the Hamming weights of $w_{ijk}$ should allow us to determine the secret key with a extremely high probability.

We can define the Hamming weight of a N-bit variable as follows, where $b_{i}$ is the $i$th bit of $x$.

\begin{equation}
\begin{aligned}
  &\mathrm{H} : \curly*{0,1}^{N} \mapsto \mathbb{N}\\
  &\ham{x} = \Sum_{i=0}^{N-1} b_{i}
\end{aligned}
\end{equation}

Since we want to compute Hamming weights it is useful to express $v_{i\,j\,k}$ and $ w_{i\,j\,k}$ in binary form as follows:
\begin{align}
  v_{i\,j\,k} &= \Sum_{n=0}^{7} c_{i\,j\,k\,n} \cdot 2^{n}\qquad\qquad c_{i\,j\,k\,n}\in\curly*{0\,1}\\
  w_{i\,j\,k} &= \Sum_{n=0}^{7} d_{i\,j\,k\,n} \cdot 2^{n}\qquad\qquad d_{i\,j\,k\,n}\in\curly*{0,1}
\end{align}

Where $c_{i\,j\,k\,n}\, d_{i\,j\,k\,n}$ represent the $n$th bit of $v_{i\,j\,k}$ and $w_{i\,j\,k}$ respectively

In order to express the xor relation described in Equation \eqref{eq:vwm}, it will be convenient to have a bit representation of $m_{l}$, where $x_{l\,n}$ will be unknowns since they represent the individual bits of the secret key.

\begin{equation}
  m_{l} = \Sum_{n=0}^{7} x_{l\,n} \cdot 2^{n}
\end{equation}

Now we can formulate the relationship of Equation \ref{eq:vwm} in a bit-wise form

\begin{equation}
  c_{i,j,(k-1),n} = d_{i,j,k,n} \oplus x_{l,n}
\end{equation}

Taking Hamming weights in both sides of said equation renders the following expression.

\begin{equation}
  \ham{v_{i,j,(k-1)}} = \ham{w_{i,j,k}  \oplus m_l}=  \Sum_{n=0}^{7} d_{i,j,k,n} \oplus x_{l,n} \label{eq:xorvw}
\end{equation}

Looking into \eqref{eq:xorvw} we can see that we will have 40 equations and 16 variables, but even and odd equations will refer to different $m_l$ and $x_{l,n}$. This is caused by the alternation in even and odd expressions. This allows us to  split this system of equations into two disjoint systems of 20 equations and 8 variables since they are completely independent of one another.

Thus, for each $j,k$ we will two systems of equations as shown in Equations \eqref{eq:system_even} and \eqref{eq:system_odd}.


\begin{equation}
\left\{
  \begin{aligned}
    \Ham{v_{0,j,(k-1)}} &= d_{0,j,k,0} \oplus x_{l,0} + d_{0,j,k,1} \oplus x_{l,1} + \ldots + d_{0,j,k,7} \oplus x_{l,7}\\
    \Ham{v_{2,j,(k-1)}} &= d_{2,j,k,0} \oplus x_{l,0} + d_{2,j,k,1} \oplus x_{l,1} + \ldots + d_{2,j,k,7} \oplus x_{l,7}\\
    \Ham{v_{4,j,(k-1)}} &= d_{4,j,k,0} \oplus x_{l,0} + d_{4,j,k,1} \oplus x_{l,1} + \ldots + d_{4,j,k,7} \oplus x_{l,7}\\
    \cdots\\
    \Ham{v_{38,j,(k-1)}} &= d_{38,j,k,0} \oplus x_{l,0} + d_{38,j,k,1} \oplus x_{l,1} + \ldots + d_{38,j,R,7} \oplus x_{l,7}\\
  \end{aligned}\right.
  \label{eq:system_even}
\end{equation}
  \vspace{.5cm}
\begin{equation}
\left\{
  \begin{aligned}
    \Ham{v_{1,j,(k-1)}} &= d_{1,j,k,0} \oplus x_{l^{\prime},0} + d_{1,j,k,1} \oplus x_{l^{\prime},1} + \ldots + d_{1,j,k,7} \oplus x_{l^{\prime},7}\\
    \Ham{v_{3,j,(k-1)}} &= d_{3,j,k,0} \oplus x_{l^{\prime},0} + d_{3,j,k,1} \oplus x_{l^{\prime},1} + \ldots + d_{3,j,k,7} \oplus x_{l^{\prime},7}\\
    \Ham{v_{5,j,(k-1)}} &= d_{5,j,k,0} \oplus x_{l^{\prime},0} + d_{5,j,k,1} \oplus x_{l^{\prime},1} + \ldots + d_{5,j,k,7} \oplus x_{l^{\prime},7}\\
    \cdots\\
    \Ham{v_{39,j,(k-1)}} &= d_{39,j,k,0} \oplus x_{l^{\prime},0} + d_{39,j,k,1} \oplus x_{l^{\prime},1} + \ldots + d_{39,j,k,7} \oplus x_{l^{\prime}\,7}\\
  \end{aligned}\right.
  \label{eq:system_odd}
\end{equation}


In general, we can solve both these systems of equations as long as we know all $d_{i,j,k,n}$ for a fixed $k$, by simply using a brute force search in the value $m_{l}$. Note that since we are performing a Simple Power Attack, careful readings will allows to the values of $\ham{v_{ij(k-1)}}$ for all $i,j,k$. The search only takes $2^{8}$ different tries with each taking at most 20 equation evaluations. We simply iterate  $m_{l}$ from $0$ to $255$ and evaluate by xoring it with $w_{i,j,k}$ and checking against $\ham{v_{i,j,(k-1)}}$ for each $i \in \curly*{0,2,\ldots 38}$. To crack $m_{l}'$ the same procedure is performed, except that this time $i$ moves in the range $i \in \curly*{1,3,\ldots 39}$. As soon as one equation is not met, we try the next possible value of $m_{l}$. Solving for a fixed $j,k$ produces two keys $m_{l}$ and $m_{l}' = m_{l+4}$ due to the even and odd scheme shown before. The algorithm employed for the brute force search is outlined in Algorithm \ref{alg:KB}.

\begin{algorithm}
    \caption{Guess Key Byte}\label{alg:KB}
    \begin{algorithmic}[1]
        \Require $\ham{v_{i,j,(k-1)}} \qquad\forall i=0,1,\ldots39$
        \Require $w_{i,j,k} \qquad\forall i=0,1,\ldots39$
        \Procedure{BreakKeyByte($\ham{v_{i,j,(k-1)}}$,$w_{i,j,k})$}{}
            \For{$m_{l} = 0$ \textbf{to} $255$}
                \State Valid $\gets$ True
                \For{$i=0$ \textbf{to} $38$ \textbf{step} 2}
                  \If{$\ham{v_{i,j,(k-1)}} \neq \ham{w_{i,j,k}  \oplus m_l}$}
                    \State Valid $\gets$ False
                    \State \textbf{break}
                  \EndIf
                \EndFor
                \If{Valid = True}
                  \State \textbf{break}
                \EndIf
            \EndFor
            \For{$m_{l+4} = 0$ \textbf{to} $255$}
                \State Valid $\gets$ True
                \For{$i=1$ \textbf{to} $39$ \textbf{step} 2}
                  \If{$\ham{v_{i,j,(k-1)}} \neq \ham{w_{i,j,k}  \oplus m_{l+4}}$}
                    \State Valid $\gets$ False
                    \State \textbf{break}
                  \EndIf
                \EndFor
                \If{Valid = True}
                  \State \textbf{break}
                \EndIf
            \EndFor
        \EndProcedure
        \State \Return $(m_{l}, m_{l+4})$
    \end{algorithmic}
\end{algorithm}

While $\ham{v_{ij(k-1)}}$ values are easy to derive from the execution of the algorithm, bit values $d_{i,j,k,n}$ will generally be unknown due to the dependence between rounds. Nevertheless, we can compute the values of $d_{i,j,k,n}$ for $k = R$. The values of $v_{i,j,R}$ are predetermined (Equation \eqref{eq:init}) and $Q_{jk}[x]$ is a known fixed permutation so we can derive the values of $w_{i,j,R}$, thus knowing all the binary variables $d_{i,j,R,n}$.
We can then solve both systems of equations for $k = R$ obtaining the 8 most significant bytes of the key.

By cracking the most significant 64 bits of the key we have obtained enough information  to calculate the values of $v$ in the next round ($v_{i,j,(R-1)}$) via equation \eqref{eq:vwm}. Using the appropriate S-boxes (Equation \eqref{eq:wQv}) we can also resolve the values of and $w$ in the next round ($w_{i,j,(R-1)}$). Since we know $\ham{v_{ij(R-2)}}$ from the measurements and we just derived $w_{i,j,(R-1)}$ we will have all $d_{i,j,R-1,n}$. So we can solve the systems of equations for $k = R - 1$ getting the following 8 bytes of the key.

By applying this scheme repeatedly we can successfully crack the whole key independently of its size with a quite narrow search space. Given $8R$ different bytes and a brute force search of at most $20 \times 2^{8}$, and with $R \leq 4$ the search space is upper bounded by $(20 \cdot 2^{8}) \cdot (8\cdot 4) < 2^{18} = 262144$.

It is important to notice that the system of equations is generally over-determined and if the Hamming values are unequivocally measured, then a compatible solution will always exist. However, there is no guarantee of the system not being under-determined. If only 7 or fewer of the 20 equations are linearly independent, multiple solutions may be possible and the described search will return the lowest one. Nevertheless, since S-boxes are designed for diffusion and we have 20 equations, the probability of this event happening is extremely unlikely. Furthermore, the modifications that are introduced later to cope with the presence of error can easily resolve this situation, so no further elaboration is needed.

The algorithm was implemented in Python and given a Hamming traces of Twofish smartcard implementation and the algorithm successfully recovered the key every time. Numerical results are shown in Section \ref{sec:results}.

\section{Attack in the presence of error} 
\label{sec:attack_in_the_presence_of_error}

Unfortunately, smart card attacks usually involve an amount of random noise overlapped with the signal. The work of Mayer-Sommer suggests that we can use correlation measures to determine Hamming weights \cite{Mayer-Sommer01smartlyanalyzing}. However, the key schedule has a significant amount of redundancy, so we can directly shield the algorithm from noise without statistical measures, simplifying the attack.

Adding noise in our measurements will mean that the values that we assumed we knew unequivocally have now added a degree of uncertainty. We can express the measured Hamming weight with added measurement error of $v_{i\,j\,k}$ as follows.
\begin{equation}
  \Ham{v_{i\,j\,k}} + \epsilon \qquad \epsilon \leftarrow \mathcal{N}(0,\upsigma^{2})
\end{equation}

Where $\epsilon$ is a random Gaussian variable with zero mean and variance $\upsigma^{2}$. Gaussian noise has been considered because even if $\epsilon$ is not a random variable, by the central limit theorem, repeated measurements over time should render a Gaussian distribution. Zero mean should be a consequence of the tuning procedure used when measuring the values, since a calibration that rendered zero mean expected error should minimize the measured error.

\subsection{Least Mean Squares} 
\label{sub:least_mean_squares}

Since the quantity measured is now real, and our algorithm worked with purely integer values, a first good step would be rounding to the nearest integer. Although not ideal, this technique will correct a significant number of our measurements. Therefore, we can define a modified Hamming function as follows, where the term $\ham{x} + \epsilon$ reflects the measure value as a whole with both the original value and the added error $\epsilon$. The operator $\curly*{\cdot} : \mathbb{R} \rightarrow \mathbb{N}$ represents the nearest integer.

\begin{equation}
  \mathrm{H}^{\ast}_{\epsilon}(x) = \curly*{\ham{x} + \epsilon }
\end{equation}

Rounding the variables to the closest integer makes the systems of equations incompatible with a extremely high probability, since they are $20 \times 8$ in size. A very common way to deal with this problem is just applying the Least Mean Squares closed form solution to get an approximation of the values. However, the systems of equations presented in Eq. \eqref{eq:system_even}, \eqref{eq:system_odd} are not linear, since they are using xors.

We can find a way of expressing the previous system as a system of linear equations. Let's start by considering that the terms $a_{ij} \oplus x_{i}$ can be decomposed as follows

\begin{equation}
  a_{ij} \oplus x_{i} = \begin{cases}
  x_{i}\phantom{ = 1 -x_{i}} \;\qquad \text{if} \quad a_{ij} =0\\
  \overline{x}_{i} = 1 -x_{i}\qquad \text{if} \quad a_{ij} =1\\
  \end{cases}
\end{equation}

By substituting the displayed expression in a arbitrary equation that follows the xor pattern shown in the systems of  equations we get:

\begin{equation}
  \begin{aligned}
  b &= \Sum_{i=0}^{7} a_{i} \oplus x_{i}\\
  b &= \Sum^{7}_{\substack{i=0 \\ a_{i} = 0}} x_{i} + \Sum^{7}_{\substack{i=0 \\ a_{i} = 1}} 1-x_{i}\\
  b &= \ham{a_{i}} + \Sum^{7}_{\substack{i=0 \\ a_{i} = 0}} x_{i} + \Sum^{7}_{\substack{i=0 \\ a_{i} = 1}} -x_{i}\\
  b - \ham{a_{i}} &= \Sum^{7}_{i=0} (-1)^{a_{i}}x_{i}
  \end{aligned}
\end{equation}

We can make use of this simplification into the previous equations and get a linear system of equations, where $a_{i\,j\,k\,n} = -2{d_{i\,j\,k\,n}}+1$. We use $-2x+1$ simply as a mapping $\curly*{0,1}\rightarrow \curly*{1,-1}$. We can show the translated formulation for a fixed $j,k$ in Equations \eqref{eq:system_even_lms} and \eqref{eq:system_odd_lms}. Since we can express the system with linear equations, we can apply the least mean squares closed form solution.

\begin{equation}
  \begin{aligned}
  A_{jk}\;\overline{x}^{*}_{l} &= \overline{h}_{j(k-1)}\\
  \overline{x}^{*}_{l} &= (A_{jk}^{\top}\;A_{jk})^{-1}A_{jk}^{\top} \;\overline{h}_{j(k-1)}\\
  \end{aligned}
\end{equation}

However, the vector $\overline{x}^{*}_{l}$ will be real valued so it needs to be mapped to $\curly*{0,1}$, so a simple mapping $\mathbb{R} \rightarrow \curly*{0,1}$ is performed.
\begin{equation}
  x_{l,n} = \max\paren*{0,\min\paren*{1,\curly*{x^{*}_{l,n}}}}
  \label{eq:mapint}
\end{equation}

Solving the equations and applying the transformation shown in Equation \eqref{eq:mapint} gives a relatively good performance for small variances $\upsigma^{2}$. The main problem arises from forward propagating errors. If we make a mistake predicting a byte of the key $m_l$ all the bytes in that row $m_{l-8},m_{l-16},\ldots$ will also be affected. Thus if we have a probability $p_1$ of making a mistake solving a specific system, the probability of making at least one mistake in that round will be $p_2 = 1-(1-p_1)^8$ since we guess 8 bytes per round. Furthermore, the probability of not making a mistake in any round out of $R-1$ rounds will go as $p_3 = (1-p_2)^{R-1} = (1-p_1)^{8(R-1)}$. Compounding these two expressions reveals the sensitivity of the algorithm to error. As the key size increases the accuracy decreases significantly.

\begin{equation}
\left\{
  \begin{aligned}
    \Hams{v_{0,j,(k-1)}} - \Ham{w_{0,j,k}} &= {a_{0,j,k,0}} \cdot x_{l,0} + {a_{0,j,k,1}} \cdot x_{l,1} + \ldots + {a_{0,j,k,7}} \cdot x_{l,7}\\
    \Hams{v_{2,j,(k-1)}} - \Ham{w_{2,j,k}} &= {a_{2,j,k,0}} \cdot x_{l,0} + {a_{2,j,k,1}} \cdot x_{l,1} + \ldots + {a_{2,j,k,7}} \cdot x_{l,7}\\
    \Hams{v_{4,j,(k-1)}} - \Ham{w_{4,j,k}} &= {a_{4,j,k,0}} \cdot x_{l,0} + {a_{4,j,k,1}} \cdot x_{l,1} + \ldots + {a_{4,j,k,7}} \cdot x_{l,7}\\
    \cdots\\
    \Hams{v_{38,j,(k-1)}} - \Ham{w_{38,j,k}} &= {a_{38,j,k,0}} \cdot x_{l,0} + {a_{38,j,k,1}} \cdot x_{l,1} + \ldots + {a_{38,j,R,7} \cdot} x_{l,7}\\
  \end{aligned}\right.
  \label{eq:system_even_lms}
\end{equation}
  \vspace{.5cm}
\begin{equation}
\left\{
  \begin{aligned}
    \Hams{v_{1,j,(k-1)}} - \Ham{w_{1,j,k}} &= {a_{1,j,k,0}} \cdot x_{l^{\prime},0} + {a_{1,j,k,1}} \cdot x_{l^{\prime},1} + \ldots + {a_{1,j,k,7}} \cdot x_{l^{\prime},7}\\
    \Hams{v_{3,j,(k-1)}} - \Ham{w_{3,j,k}} &= {a_{3,j,k,0}} \cdot x_{l^{\prime},0} + {a_{3,j,k,1}} \cdot x_{l^{\prime},1} + \ldots + {a_{3,j,k,7}} \cdot x_{l^{\prime},7}\\
    \Hams{v_{5,j,(k-1)}} - \Ham{w_{5,j,k}} &= {a_{5,j,k,0}} \cdot x_{l^{\prime},0} + {a_{5,j,k,1}} \cdot x_{l^{\prime},1} + \ldots + {a_{5,j,k,7}} \cdot x_{l^{\prime},7}\\
    \cdots\\
    \Hams{v_{39,j,(k-1)}} - \Ham{w_{39,j,k}} &= {a_{39,j,k,0}} \cdot x_{l^{\prime},0} + {a_{39,j,k,1}} \cdot x_{l^{\prime},1} + \ldots + {a_{39,j,k,7}} \cdot x_{l^{\prime},7}\\
  \end{aligned}\right.
  \label{eq:system_odd_lms}
\end{equation}


\subsection{Minimizing Hamming difference} 
\label{sub:minimizing_Hamming_difference}

In order to circumvent the problem outlined in the previous section, we can use the high redundancy built into the system to correct the mistakes made by the least mean squares approximation.

When we make a mistake solving for $m_l$ it is because one or more of the bits $x_{l,n}$ are incorrect. The most common scenario is one incorrect bit, then two incorrect bits and so on and so forth. Thus, we would like to try to correct the possible the errors in order of increasing complexity.

We can solve this issue by ordering the set of integers by Hamming weight as follows

\begin{equation}
  H_M = \curly*{0,1,2,4,8,16,32,64,128,3,5,9,\ldots,254,255}
\end{equation}

Given this order, we can xor these masks to toggle one or several bits of the estimated key $m_l$ producing the set of candidate keys ordered by Hamming distance to the original estimate.
\begin{equation}
  M_l = \curly*{m_l \oplus h_M : \quad \forall h_M \in H_M}
\end{equation}
However we need a measure of how good is the fit of an arbitrary $m'_l \in M_l$ with respect to other candidates in the set. A good approach is to minimize the sum of Hamming distances of the predicted values $\Ham{w_{i,j,k} \oplus m'_{l}}$ after xoring with $m'_l$ and the rounded measured Hamming weights $\Hams{v_{i,j,(k-1)}}$. However, this can result in over-fitting the key to this particular measure by getting a mask with a unnecessary large Hamming weight just because it renders the largest value for this measure. To compensate this effect we also try to minimize the sum of Hamming distances at the output of the substitution boxes, comparing the predicted values $\Ham{Q_{jk}\paren*{w_{i,j,k} \oplus m'_{l}}}$ with the rounded measured values $\Hams{w_{i,j,(k-1)}}$. Until now we only needed the trace of Hamming weights of $\boldsymbol{V}$, but to implement this correction the trace of $\boldsymbol{W}$ will also be needed.

\begin{equation*}
  \label{eq:objfun}
  m_l^{*} = \underset{m'_l \in M_l}{\text{argmin}} \curly*{\Sum_{i=0}^{18} \abs*{ \Ham{w_{i,j,k} \oplus m'_{l}} - \Hams{v_{i,j,(k-1)}} } + \Sum_{i=0}^{18} \abs*{ \Ham{Q_{jk}\paren*{w_{i,j,k} \oplus m'_{l}}} - \Hams{w_{i,j,(k-1)}} }}
\end{equation*}

The algorithm was implemented and the accuracy of the algorithm was significantly improved under the presence of error, even with high variance.

The main issue with this approach is caused by how computationally expensive becomes optimizing this function. In order to simplify it, we can restrict these masks to have a maximum Hamming weight. For example, if we only consider the mask with weight smaller or equal to two we reduce the space of possible masks from 255 to 37. Therefore, the subset of masks with Hamming weights smaller than a threshold $\tau$ is defined as follows:

\begin{equation}
  \widetilde{M}_l^{\tau} = \curly*{m_l \oplus h_M : \quad \Ham{h_M} < \tau \quad \forall h_M \in H_M  }
\end{equation}
The size of this set does not grow linearly, due to the combinatorial terms. In general increasing $\tau$ means that we can resolve larger amounts of error, but the procedure becomes more computationally expensive as we show in the numerical results in the next section.
\begin{equation}
  \abs*{\widetilde{M}_l^{\tau}} = \Sum_{n=0}^{\tau} \binom 8n
\end{equation}


\ifEXTENDED
\subsection{Multiple Readings} 
\label{sub:multiple_readings}

The analysis so far was done considering just a single reading of the execution of the algorithm. Nevertheless, in real life it is not far-fetched to obtain multiple readings that use the same secret key. In general, the most difficult part consists on identifying which reading is associated with each secret key. However, given an algorithm as the one outlined in Section \ref{sub:minimizing_Hamming_difference}, we can easily cluster the readings using some kind of similarity measure between the sequences. Direct statistical analysis such as Pearson correlation would not take into account the order of the sequence which is of extreme relevance in this case. Therefore, we can represent the key estimates as elements of $\mathbb{Z}^{N}_{256} \quad N = \curly*{16,24,32}$ and compute the euclidean distance between them. The cardinality of $\mathbb{Z}^{N}_{256}$ is the same as $\mathbb{Z}^{8N}_{2}$, so it is extremely unlikely to randomly sample two different keys which have a small euclidean distance.

\begin{equation}
  \norm*{K-K'}_{2} = \sqrt{\sum_{i=1}^{256} (K_{i}-K_{i}')^{2}}
  \qquad K,K' \in \mathbb{Z}^{N}_{256}
\end{equation}

In a real world scenario, our eavesdropper device would get a set of power readings. For each one of them we would get a different key estimate. We could then use a threshold radius tuned based on simulation results to easily identify key estimates with small distance values. Thus, we would get a series of clusters, one for each different key that the eavesdropper would have observed.

Moreover, as we have previously stated, the estimates for the individual bytes of the key for each round are independent of each other. Namely, a byte $m_i$ will only depend on bytes from previous rounds $m_{i+8}, m_{i+16},\ldots$. We can see that when making a mistake in $m_i$, it is going to propagate through the algorithm and affect the sequence $m_{i},m_{i-8},\ldots$. As we saw in Figure \ref{fig:keydetail}, there are 8 different sets of rows in the key schedule, so the probability of making a mistake in the same byte in two different readings is small. Consequently, the probability of making a mistake in the same byte using both Least Mean Squares plus Hamming mask correction and outputting the same result for a particular byte is extremely slim.

Therefore, if we had access to multiple power traces associated with different readings we would be able to correct for these simple mistakes by just performing a majority vote. Our approach treated the multiple readings independently and to get the final estimate of the key just took the most common value along the predicted bytes for that particular position. As long as we have at least two predicted keys that share the same byte estimate for each byte, we can be quite confident that the predicted key will be the original one.

Finally, if for a specific byte the majority vote does not output any value, namely that all byte estimates are equally likely, we could still compare which was the error for each Least Min Squares estimate, what was the weight of the hamming mask employed in the correction and which error did the objective function \eqref{eq:objfun} achieve. Comparing these values, we would be able to tell which byte is the most likely one.

\fi


\section{Results} 
\label{sec:results}

In this section the numerical results for accuracies and runtimes are shown for the different algorithms outlined in Section \ref{sec:twofish_key_schedule_power_analysis_attack}. All the implementation was performed in Python 2.7.11 (64-bit) and run using a single core at \SI{2.6}{\giga\Hz}.

\subsection{Simple Power Attack} 
\label{sub:simple_power_attack}

The results for the naive approach (the one that does not take measurement error into consideration) are shown in Table \ref{tab:noerror}. A sample of $1000$ random runs was used to test the performance of the algorithm. As we can see, the algorithm predicts correctly the key every time and the runtimes are in the domain of tenths of seconds.

\begin{table}[ht]
    \centering
    \begin{tabular}{lrr}
        {Key Size} & {Accuracy} & {Avg. Runtime}\\
        \midrule
        128 & 100\% & \SI{3.75}{\ms}\\
        192 & 100\% & \SI{5.7}{\ms}\\
        256 & 100\% & \SI{7.39}{\ms}\\
    \end{tabular}
    \caption{Performance of the algorithm for $1000$ random keys per key size.}
    \label{tab:noerror}
\end{table}


\subsection{Error correction} 
\label{sub:error_correction}

In order to evaluate the performance of the error correction algorithm we need to compute the performance for several values of the Standard deviation $\upsigma$. Furthermore, as we stated previously, varying the maximum Hamming weight $\tau$ of the masks used will result in a trade-off between accuracy and runtime.

\begin{figure}[htp]
    \centering
    \begin{subfigure}{0.48\textwidth}
        \centering
        \includegraphics[width=\textwidth]{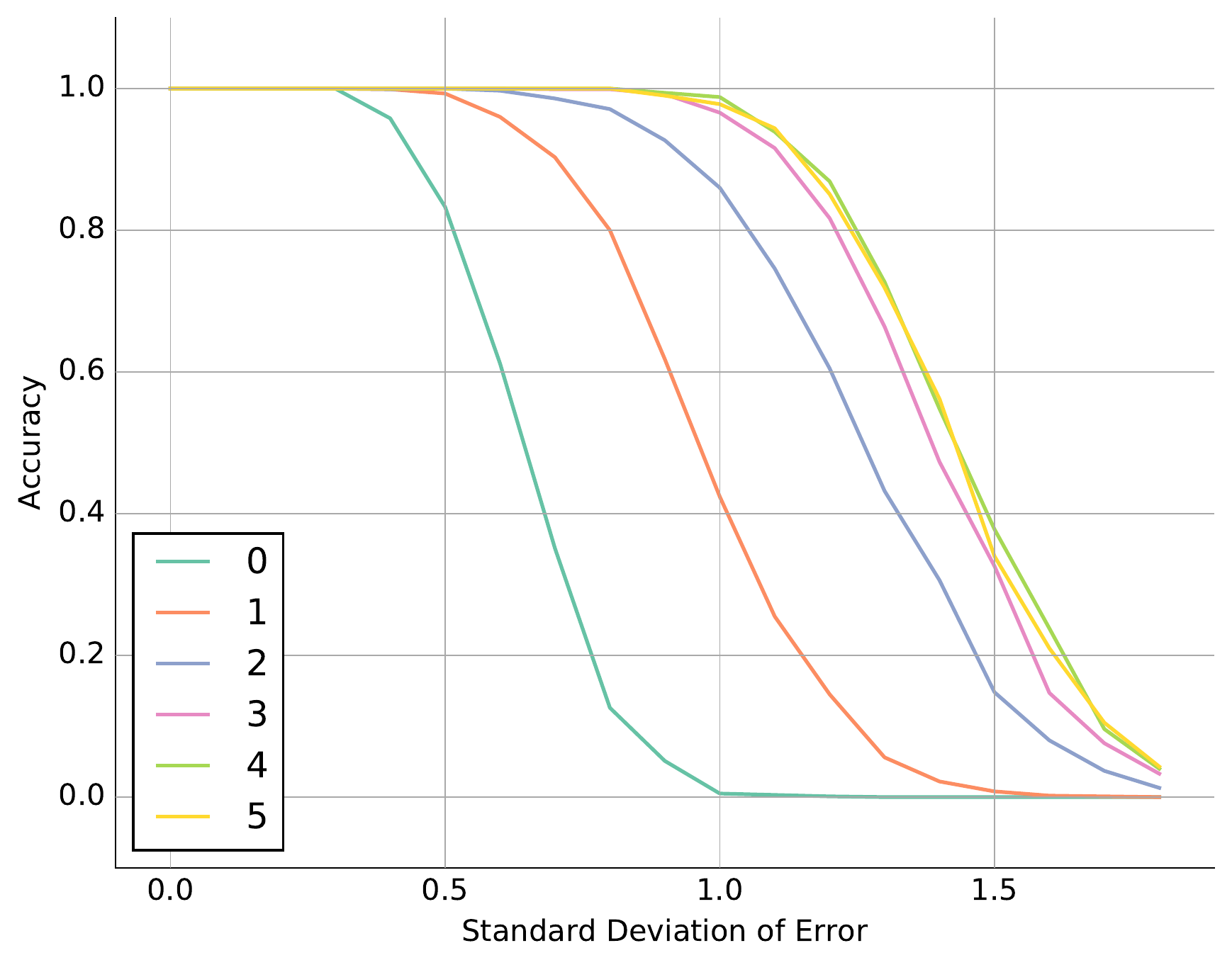}
        \caption{Average accuracies for 128-bit key}
    \end{subfigure}
    \hspace{.35cm}
    \begin{subfigure}{0.48\textwidth}
        \centering
        \includegraphics[width=\textwidth]{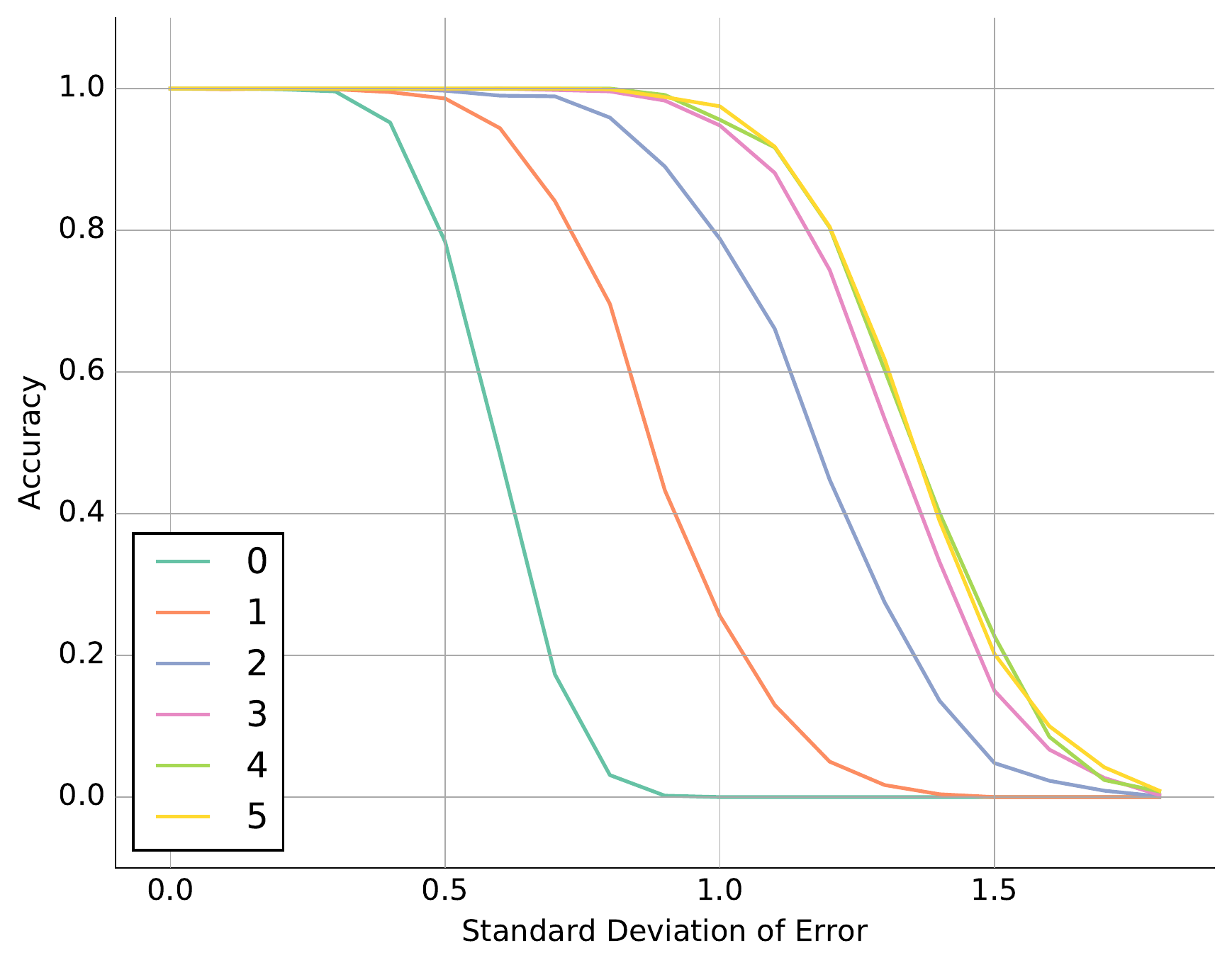}
        \caption{Average accuracies for 192-bit key}
    \end{subfigure}%

    \vspace{1cm}
    \begin{subfigure}{0.48\textwidth}
        \centering
        \includegraphics[width=\textwidth]{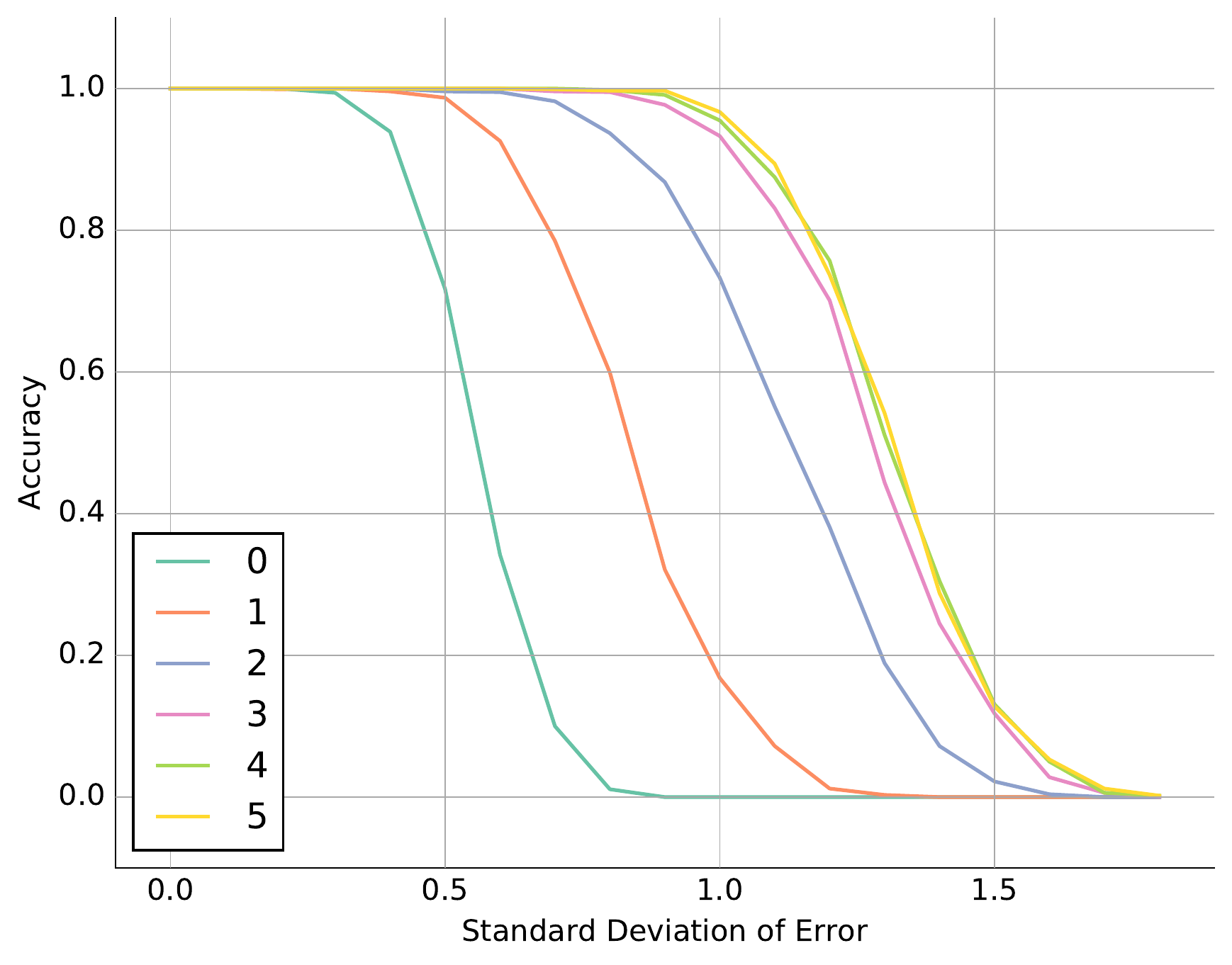}
        \caption{Average accuracies for 256-bit key}
    \end{subfigure}
    \hspace{.35cm}
    \begin{subfigure}{0.48\textwidth}
        \centering
        \includegraphics[width=\textwidth]{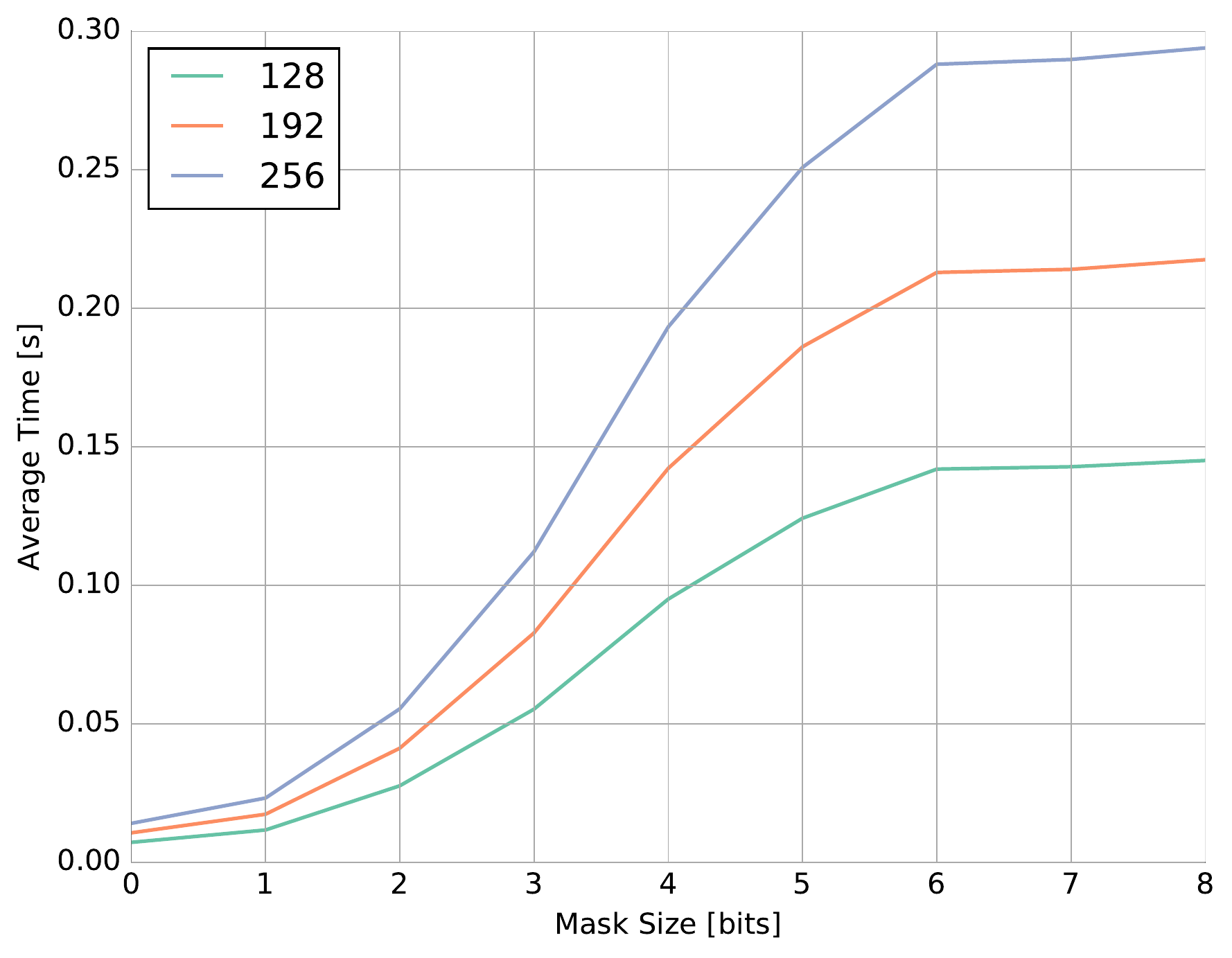}
        \caption{Average Runtimes for various mask sizes}
    \end{subfigure}
    \vspace{.25cm}
    \caption{Results of simulation of the LMS algorithm with correction}
    \label{fig:plots}
\end{figure}

Therefore, a simulation of random keys was performed for values of $\upsigma = 0.1,0.2,\ldots,2.0$ and $\tau = 0,1,\ldots 8$. Note that for $\tau = 0$ the algorithm does not perform any sort of optimization and is equivalent to the one described in Section \ref{sub:least_mean_squares}. A thousand random runs of the algorithm per combination of $\upsigma$ and $\tau$ were performed. Table \ref{tab:error_128} shows average accuracies and runtimes for several $\upsigma$ and $\tau$. Average runtime $t$ does not depend on $\upsigma$ so it is aggregated in the last row. We can also visualize the results of the simulation in Figure \ref{fig:plots}.

From the values shown on the Table we can draw several conclusions. It is easy to see an almost diagonal pattern, since higher values of $\upsigma$ decrease the accuracy and higher values of $\tau$ increase it. However, for values of $\tau > 3$ the improvement is negligible, whereas the runtime experiments a threefold increase. The times have multiplied from our original brute force search algorithm. For $\tau =0$ (no correction performed) the runtime is doubled and the accuracy is poor for values $\upsigma > 0.6$. Furthermore, employing $\tau = 3$ as factor correction increases the runtime by a factor of 8. Nevertheless, increasing one order of magnitude the runtime almost completely shields the algorithm from Gaussian errors with standard deviation $\upsigma < 1.0$, which is a more than considerable amount of noise.

\begin{table}[ht]
    \centering
    \begin{tabular}{rrrrrrrrrr}
        $\upsigma \setminus \tau$ & \multicolumn{1}{c}{0} &\multicolumn{1}{c}{1} &\multicolumn{1}{c}{2} &\multicolumn{1}{c}{3} &\multicolumn{1}{c}{4} &\multicolumn{1}{c}{5} &\multicolumn{1}{c}{6} &\multicolumn{1}{c}{7} &\multicolumn{1}{c}{8}\\
         \midrule
        0.0 & 100.0  &100.0  &100.0  &100.0  &100.0  &100.0  &100.0  &100.0  &100.0 \\
        0.2 & 100.0  &100.0  &100.0  &100.0  &100.0  &100.0  &100.0  &100.0  &100.0 \\
        0.4 & 97.5  &99.8  &100.0  &100.0  &100.0  &100.0  &100.0  &100.0  &100.0 \\
        0.6 & 63.2  &96.7  &99.7  &100.0  &100.0  &100.0  &100.0  &100.0  &100.0 \\
        0.8 & 12.4  &77.3  &96.9  &99.8  &99.9  &100.0  &100.0  &100.0  &99.8 \\
        1.0 & 1.4  &42.1  &87.2  &97.3  &98.1  &98.0  &97.7  &98.7  &98.4 \\
        1.2 & 0.0  &12.8  &62.4  &83.2  &84.1  &84.9  &83.7  &85.5  &86.2 \\
        1.4 & 0.0  &3.1  &27.0  &52.6  &56.1  &56.0  &56.3  &53.5  &54.8 \\
        1.6 & 0.0  &0.4  &8.4  &15.9  &21.7  &21.3  &19.6  &23.4  &22.3 \\
        1.8 & 0.0  &0.0  &2.3  &3.6  &4.2  &5.4  &4.9  &4.8  &4.7 \\
        2.0 & 0.0  &0.0  &0.1  &0.5  &0.6  &0.6  &0.3  &0.1  &0.7 \\
        \midrule
         $t$(ms) &7.27 &11.52 &25.94 &55.36 &91.46 &121.11 &148.50 &150.94 &142.14\\
    \end{tabular}
    \caption{Performance of the algorithm for 128-bit key and several Standard deviations and mask sizes}
    \label{tab:error_128}
\end{table}

Accompanying graphics depicting the performance for different key sizes, amounts of noise and mask sizes are included in Figure \ref{fig:plots}. From these Figures shown below we can make a number of remarks. First, the results agree with the conclusions form the analysis of Table \ref{tab:error_128}. Furthermore, since here can compare the performance for different key sizes we can observe that the higher the key size, the smaller the error tolerance of the algorithm. This was expected since as we commented previously, there exists a forward propagation of errors, so bigger key sizes will decrease the probability of cracking the key. In the figures we can also see that for $\tau \geq 3$ the improvement is almost negligible, so large values of $\tau$ have been omitted for clarity.

\ifEXTENDED
\subsection{Multiple Readings} 
\label{sub:multiple_readings2}

As we outlined in Section \ref{sub:multiple_readings} if we observe multiple power traces from the same key we will be able to improve the error correction results for the algorithm. To analyze this behavior we performed a simulation that employed up to 5 readings per secret key. If the algorithm was able to obtain a majority vote for each byte of the key without ties it would return that key. Otherwise, it would execute a new simulation with that key up to a limit of 5 readings. We decided to use 5 readings as an upper limit since it gave significant results and it is small enough to be observed in a real world scenario.

Similarly we run a thousand random runs for varying values of the parameters $\upsigma$ and $\tau$ we run one thousand random simulations. Results can be seen in Table \ref{tab:error_128_avg} for 128-bit key. We find the same pattern as in the previous table, except that in this case error tolerances have been significantly improved. As we saw before results are almost identical for $\tau \geq 4$. Accuracies have improved for all combinations of $\upsigma, \tau$ at the expense of times increasing by an average of a factor of three.

 \begin{table}[ht]
     \centering
     \begin{tabular}{rrrrrrrrrr}
         $\upsigma \setminus \tau$ & \multicolumn{1}{c}{0} &\multicolumn{1}{c}{1} &\multicolumn{1}{c}{2} &\multicolumn{1}{c}{3} &\multicolumn{1}{c}{4} &\multicolumn{1}{c}{5} &\multicolumn{1}{c}{6} &\multicolumn{1}{c}{7} &\multicolumn{1}{c}{8}\\
          \midrule
         0.0 & 100.0  &100.0  &100.0  &100.0  &100.0  &100.0  &100.0  &100.0  &100.0 \\
         0.2 & 100.0  &100.0  &100.0  &100.0  &100.0  &100.0  &100.0  &100.0  &100.0 \\
         0.4 & 100.0  &100.0  &100.0  &100.0  &100.0  &100.0  &100.0  &100.0  &100.0 \\
         0.6 & 98.6  &100.0  &100.0  &100.0  &100.0  &100.0  &100.0  &100.0  &100.0 \\
         0.8 & 84.7  &99.5  &100.0  &100.0  &100.0  &100.0  &100.0  &100.0  &100.0 \\
         1.0 & 28.1  &98.3  &99.9  &100.0  &100.0  &100.0  &100.0  &100.0  &100.0 \\
         1.2 & 1.3  &88.7  &99.4  &99.9  &99.8  &99.9  &100.0  &99.9  &100.0 \\
         1.4 & 0.0  &57.1  &96.2  &99.3  &99.0  &99.3  &99.4  &99.8  &99.7 \\
         1.6 & 0.0  &18.6  &81.2  &93.0  &93.7  &94.8  &95.6  &92.3  &93.3 \\
         1.8 & 0.0  &2.4  &42.8  &67.0  &70.1  &72.1  &69.0  &69.1  &68.6 \\
         2.0 & 0.0  &0.1  &9.0  &23.1  &24.5  &26.9  &28.2  &27.3  &27.3 \\
         \midrule
          $t$(ms) &27.92 &40.23 &77.30 &157.27 &252.05 &342.18 &381.46 &399.85 &413.72\\
     \end{tabular}
     \caption{Performance of the algorithm for 128-bit key and with multiple readings per key}
     \label{tab:error_128_avg}
 \end{table}

Accompanying graphics are also included for this case, depicted in Figure \ref{fig:plots_avg}. From the Figures we can see that for all key sizes $N = \curly*{128, 192, 256}$ we have improved the error tolerance from $\upsigma \approx 1.0$ to $\upsigma \approx 1.5$. The main difference when comparing these Figures to the ones  in the previous section is the slope of the curve in the transition. Without multiple readings, the slope was fairly smooth, similar to that of a sigmoid function. Nevertheless, in the new graphics we observe that the transition curve is more abrupt, specially for 192 and 256 bit keys. This makes sense since we are introducing more layers of complexity to the error correction. The corrections make the algorithm more sensible to errors when it is unable to compensate for the amount of noise being added into the readings. Finally, for the time representation we have changed the grouping by mask size to standard deviation since now, depending on the amount of error the algorithm will do between two and five executions to determine the final key. Thus, we can see that for $\upsigma \lesssim 1.0$ two executions are enough to correct all of the errors but for bigger values of the noise we start needing more executions to achieve a clear majority vote. Transition happens at $\upsigma \approx 1.5$, which agrees with the accuracy results obtained in the rest of the Figures.
\begin{figure}[htp]
    \centering
    \begin{subfigure}{0.48\textwidth}
        \centering
        \includegraphics[width=\textwidth]{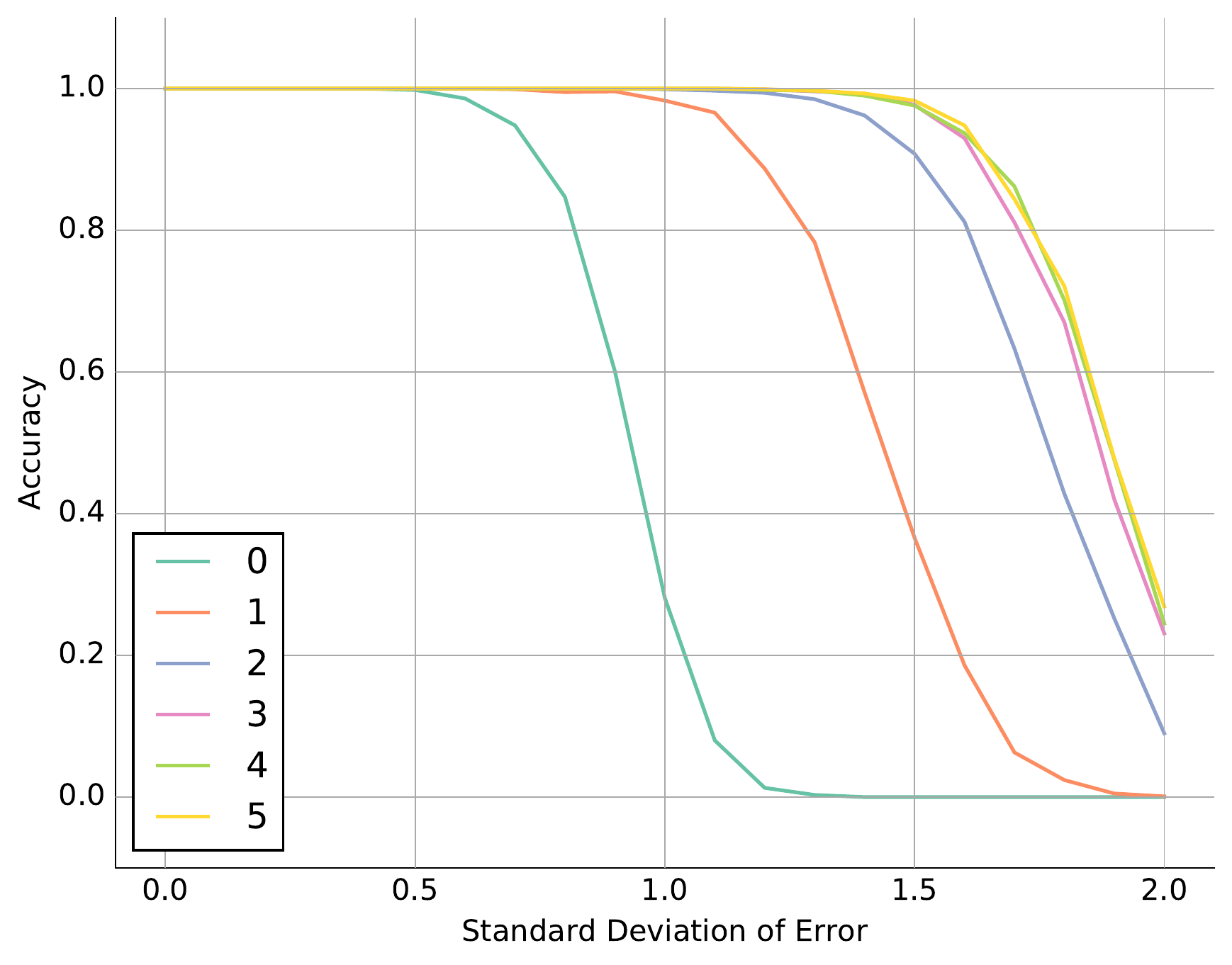}
        \caption{Average accuracies for 128-bit key}
    \end{subfigure}
    \hspace{.35cm}
    \begin{subfigure}{0.48\textwidth}
        \centering
        \includegraphics[width=\textwidth]{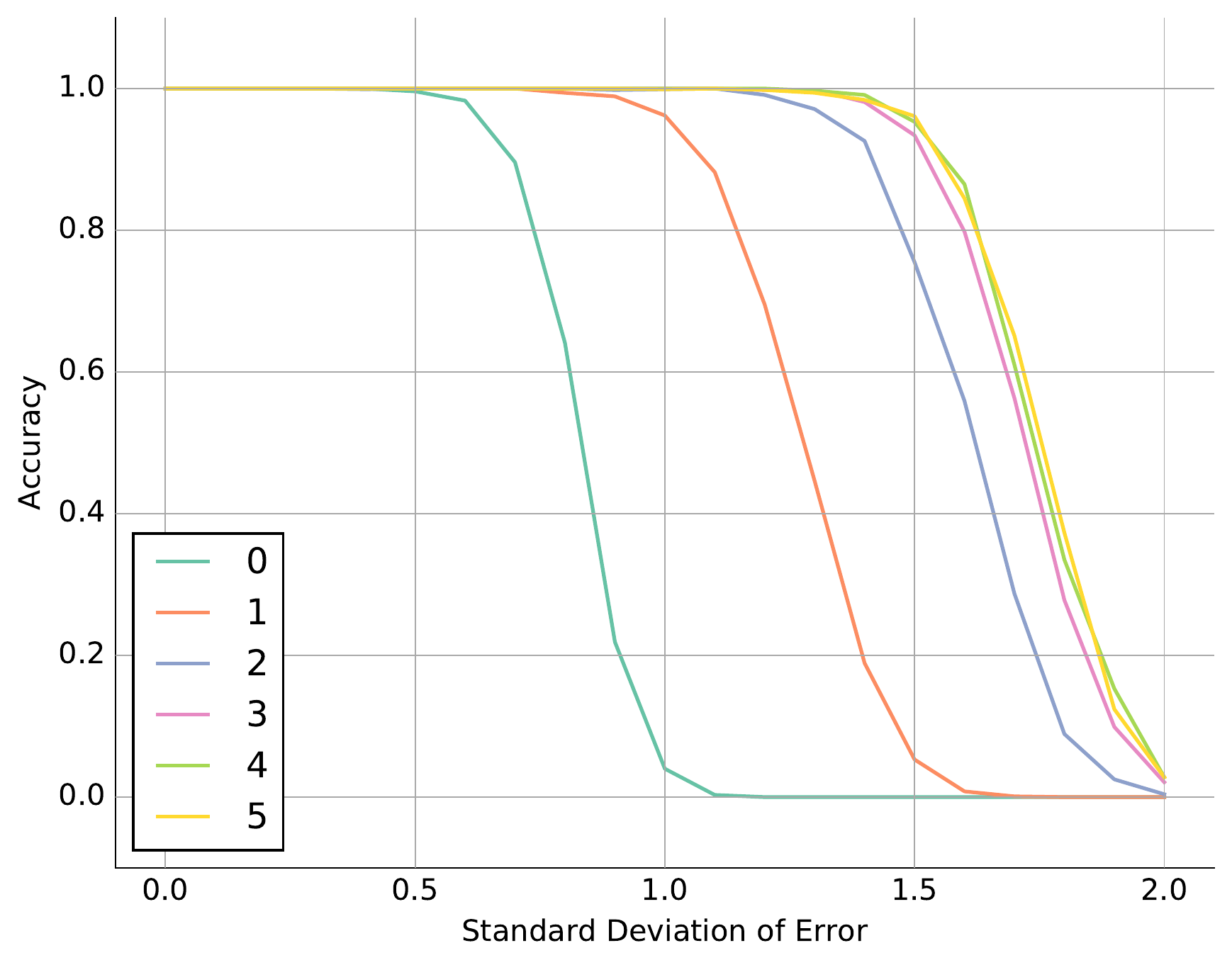}
        \caption{Average accuracies for 192-bit key}
    \end{subfigure}%

    \vspace{1cm}
    \begin{subfigure}{0.48\textwidth}
        \centering
        \includegraphics[width=\textwidth]{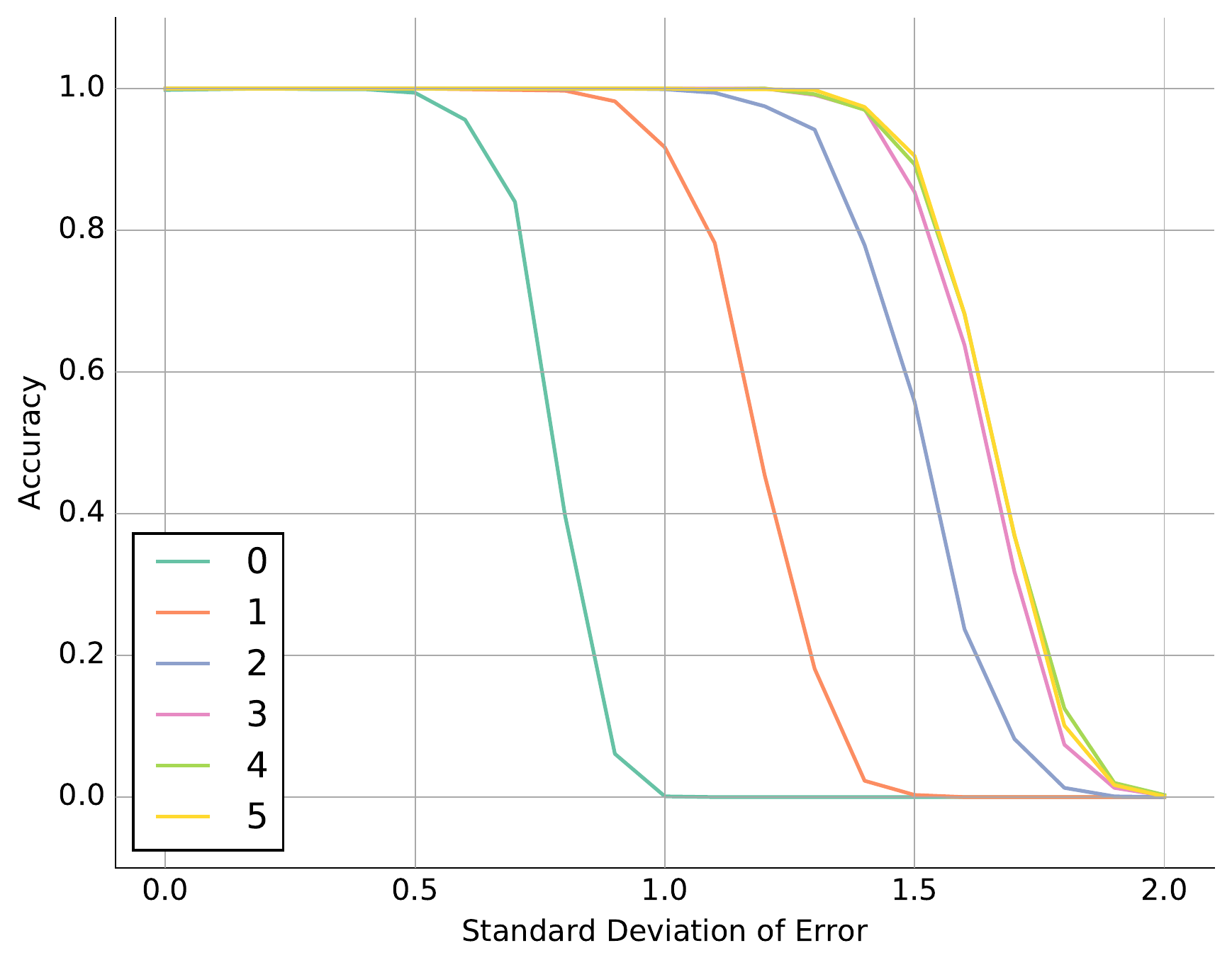}
        \caption{Average accuracies for 256-bit key}
    \end{subfigure}
    \hspace{.35cm}
    \begin{subfigure}{0.48\textwidth}
        \centering
        \includegraphics[width=\textwidth]{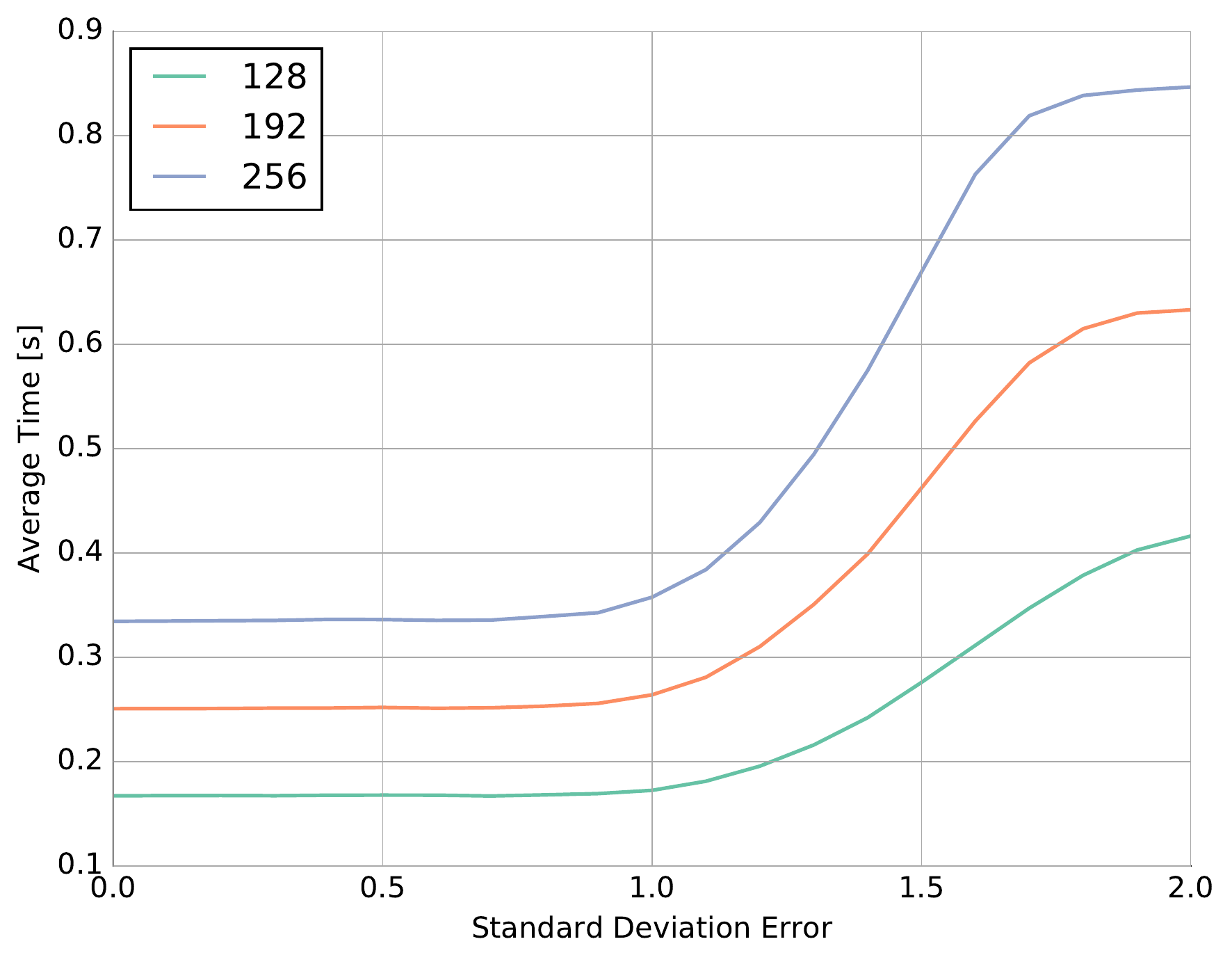}
        \caption{Average Runtimes for various noise levels}
    \end{subfigure}
    \vspace{.25cm}
    \caption{Results of simulation with multiple readings per key}
    \label{fig:plots_avg}
\end{figure}

\fi

\section{Discussion} 
\label{sec:discussion}

A main caveat of Simple Power Attacks is that even they work in theory, they usually need a perfect Hamming trace of the execution. However, in practice we tend to experience a number of inconveniences, such as noise, uncorrelated measurements with the execution of the program and even schemes that masquerade any useful values. Therefore, a good SPA attack will not only work in theory but will also perform reasonably well under these handicaps.

Although we did not comment it in the previous sections, obtaining a value correlated to the Hamming values has been proved to be possible to perform as shown in \cite{Mayer-Sommer01smartlyanalyzing}. The Twofish implementation for several languages is publicly available \cite{TSC}. Therefore, if we can run the algorithm step by step we will be able to establish the specific locations where the algorithm computes values whose Hamming weights are required for the described attack. Careful timing and some statistical analysis should resolve in the sought values when performing the attack. \cite{Mangard:2007:PAA:1208234}

Among the implementations are a couple of 8-bit assembly languages (Motorola 6800 and Zilog Z80) as well as more complex implementations in 32-bit and 64-bit environments. The attack works directly on both 8-bit implementations since is inherently designed for 8-bit Hamming weights. Nevertheless, examining closely the x86 Assembly and C implementations we can see that the S-box is performed as a look-up table operation. Thus, words are shifted into bytes to perform this look-up so the values needed for the attack can also be found in these implementations. Twofish's description of the $h$ function with such an arbitrary arrangement of S-boxes makes extremely difficult and inefficient to program an implementation that does not get the individual 8-bit values to perform the S-box substitution. Thus, as long as we can get a reliable power analysis trace, any Twofish implementation will most likely be vulnerable to the attack presented in this paper.

Most systems vulnerable to SPAs will not give a perfect power analysis trace, so we should expect a considerable amount of noise added to our measurements. As we saw in Sections \ref{sec:attack_in_the_presence_of_error} and \ref{sec:results}, modifying the algorithm to perform under error was possible. Moreover, as the results displayed, the algorithm is quite robust to Gaussian noise.

Comparing the accuracy graphics with the time plot, we can conclude that the best trade-off accuracy-time happens at $\tau = 3$. This is probably linked to the fact that even with a high standard deviation is hard to have more than 3 bits consistently changed for the twenty bytes that we sample in the trace. Moreover, higher values of $\tau$ are not able to account for higher values of $\upsigma$. Namely, for $\upsigma \lesssim 1.0$, we can just increase the threshold weight $\tau$ and we will manage to get $100\%$ accuracy. However, for values $\upsigma \gtrsim 1.0$, increasing $\tau$ does not increase accuracy and in fact, for $\upsigma = 2.0$ the algorithm almost always makes a mistake in at least one byte of the key.


\section{Conclusion} 
\label{sec:conclusion}

We have shown that the Twofish Encryption Standard is susceptible to a Simple Power Attack that is based solely in the Hamming weights of the Key Schedule Computation. The algorithm successfully obtains the value of the secret with just one run of the algorithm and in the presence of a relatively large amount of noise. Moreover, we have seen that the attack threatens not only 8-bit implementations but any Twofish implementation since the architecture of the cipher forces the key schedule to explicitly compute the byte values that are needed for the described attack. Finally, the worst runtime of the algorithm is under one second, so the attack is not only feasible but also efficient.

\subsection{Future Work} 
\label{sub:future_work}

As we introduced earlier, previous research has shown that both Rijndael \cite{VanLaven2005SideCA} and Serpent \cite{Compton2009ASP} key schedules were susceptible to Simple Power Attacks. Twofish brings to three the list of AES finalist with known SPA attacks. Thus, it would be interesting to analyze the possibility of having similar SPA in the MARS and RC6 key schedules. If they were to be susceptible to similar SPAs it would set an interesting precedent for the next generation of cryptographic standards to consider. Embedded systems are becoming more and more widespread and with the introduction of frameworks such as  \emph{the Internet of Things}, confidentiality and integrity of the information acquired and transmitted by these devices will be an important concern.

Moreover, regarding the followed approach a number of considerations for further extensions can be made. The first remark is that we did not use the entirety of the key related information. As we outlined in Section \ref{sec:twofish_block_cipher}, each round of the algorithm uses the words $S_i$, which are derived applying a Reed Solomon Transformation to the secret key. From the Hamming trace of the matrix calculations, a more refined search could be done to correct for errors. Furthermore, in the optimization function we only considered the input and output of a single stage of substitution. A more global approach could consider the input and output of the subsequent stages and search for the minimum of those and backtrack as necessary. Nevertheless, since the runtime would increase in combinatorial way, a smart algorithm would be needed to just consider the most likely cases and prune said search.

Finally, it would be interesting to gather real world data and perform the attack on a real system in order to consider for the statistical inference problems related to find the appropriate values in trace as well as to confirm the conclusions derived in this work.


\nocite{*}
\bibliographystyle{alpha}
\bibliography{mybib}


\end{document}